\documentstyle[preprint,aps,eqsecnum]{revtex}
\begin{document}
\draft
\title{ Fermionic Chern-Simons theory for the
Fractional Quantum Hall Effect in Bilayers}
\author{Ana Lopez and Eduardo Fradkin}
\address{Department of Physics, University of Illinois at
Urbana-\-Champaign , 1110 West Green Street,Urbana, IL 61801-3080}
\bigskip

\maketitle

\begin{abstract}
We generalize the fermion Chern-Simons
theory for the Fractional Hall Effect (FQHE) which we developed before, to the
case of bilayer systems.
We study the complete dynamic response of these systems and predict
the experimentally accessible optical properties.
In general, for the so called $(m, m, n)$ states, we find that the
spectrum of collective excitations has a gap, and the wave
function has the Jastrow-Slater form, with the exponents determined by the
coefficients $m$, and $n$.
We also find that the  $(m,m,m)$ states, {\it i.~e.~}, those states
whose filling fraction is $1\over m$, have a gapless mode which may be
related with the spontaneous appearance of the interlayer coherence.
Our results also indicate that the gapless mode makes a contribution to
the wave function of the $(m,m,m)$ states analogous to the phonon
contribution to the wave function of superfluid $\rm{He}_4$.
We calculate the Hall conductance, and the charge and statistics of the
quasiparticles. We also present an $SU(2)$ generalization of this
theory relevant to spin unpolarized or partially polarized single
layers.
\end{abstract}

\bigskip


\narrowtext

\section{Introduction}
\label{sec:intro}

The Quantum Hall Effect is observed in two dimensional electron systems (2DES)
in the presence of strong perpendicular magnetic fields, at very low
temperatures.
This effect is characterized by the existence of an energy gap between
the ground state and the lowest excited state. In the case of the Integer
Quantum Hall Effect (IQHE) the energy gap is the Landau level spacing
produced by the external magnetic field at integer filling factors. In the
Fractional Quantum Hall Effect (FQHE) the energy gap appears as a result of
the interparticle correlations due to the strong interactions between the
electrons.

If one allows for the presence of new degrees of freedom, a richer variety
of states can be found. The two obvious possibilities that one can consider
are systems in which the electronic spin is not frozen by the Zeeman energy,
and systems in which two or more layers of 2DES are coupled together.
For instance, the experimentally observed $\nu ={5\over 2}$ state \cite{w52},
has been explained theoretically by Haldane and Rezayi \cite{hr} using the
fact that the system is not spin-polarized.

Due to continuing advances in material-growth techniques, it has been possible
to fabricate high-quality multiple 2DE layers in close proximity. In these
systems the layer index is the new degree of freedom, and the interplay
between the intralayer and the interlayer Coulomb interactions gives rise to
very interesting physics. In particular, this competition can explain
\cite{mpb} the experimental observation \cite{e2} of the destruction or
weakening of the IQHE at odd filling fractions.
 Another interesting case is the one of the $\nu = {1\over 2}$ state.
In single-layer systems, even though many transport anomalies have been
reported, there is no evidence of FQHE. On the other hand, this is a
well observed \cite{unmed} FQHE state in double-layer systems.

Motivated by the fact that very interesting physics can be found in these
2DES if one considers new degrees of freedom, we study double-layer FQHE
systems. Our formalism can also be extended to the study of spin
non-polarized systems.

There are two energy scales that play a very important role in this problem.
One is the potential energy between the electrons in different layers, and
the other one is the tunneling amplitude between layers.
We only consider the case in which the
tunneling between the layers may be neglected, and both layers are identical.
Therefore, the number of particles
in each layer is conserved, and the collective modes corresponding to
in phase and out of phase density oscillations are decoupled.

We generalize the fermionic Chern-Simons field
theory developed in reference \cite{l1}. The generalization is straightforward.
We consider a theory in which the electrons are coupled to both
the electromagnetic field, and to the Chern-Simons gauge fields (two in
this case, one for each layer). We show that
this theory is equivalent to the standard system in which the Chern-Simons
fields are absent, provided that the coefficient of the Chern-Simons action
is such that the electrons are attached to an even number of fluxes of the
gauge field in their own layer, and to an arbitrary number of fluxes of the
gauge field in the opposite layer. In this form, the theory has a $U(1)
\otimes U(1)$ gauge invariance. We obtain the same action as the one
derived by Wen and Zee in their matrix formulation of topological fluids
\cite{wen}.

In this paper, we study the liquid-like solution of the
semiclassical approximation to this theory. We can describe a
large class of states which are characterized by filling fractions
in each layer given by
\begin{eqnarray}
\nu _{1} &=& {{n-(\pm {1\over p_{2}} + 2 s_{2})} \over
              { n^{2} - (\pm {1\over p_{1}} + 2 s_{1})(\pm {1\over p_{2}}
+ 2 s_{2})}}
\nonumber \\
\nu _{2} &=& {{n-(\pm {1\over p_{1}} + 2 s_{1})} \over
              { n^{2} - (\pm {1\over p_{1}} + 2 s_{1})(\pm {1\over p_{2}}
+ 2 s_{2})}}
\label{eq:fillf}
\end{eqnarray}
where $p_{1}$, $p_{2}$, $s_{1}$, $s_{2}$, and $n$ are integers.
This includes the so called
($m_{1}, m_{2},n$) with filling fractions
$\nu={\frac{2n-m_1-m_2}{n^2-m_1m_2}}$, and the ($m, m, m$) states, with
filling fractions $\nu={\frac{1}{m}}$.
We calculate the electromagnetic response functions and find the spectrum
of collective excitations. We find that for the ($m_{1}, m_{2},n$) states the
{\it in phase} as well as the {\it out of phase} collective
excitations are gapped.
On the other hand, for the ($m, m, m$) states there is an {\it out of phase}
gapless mode which, in the absence of tunneling between layers, indicates the
spontaneous  breaking of the U(1) symmetry associated with the conservation
of the relative number of particles.

We show that, already at the semiclassical level of our approach, the
density correlation functions
saturate the $f$-sum rules, associated with the two separated conservation
laws,
{\it i.~e.~}, the number of particles on each plane is separately
conserved.
Using this property, we can derive the universal form of the absolute
value squared of the ground state wave function at long distances and
in the thermodynamic limit. For the ($m, m, n$) states the wave function that
we find
has the Jastrow form predicted by Halperin \cite{halperin}. For the
($m, m, m$) states, we find that there is an additional factor which
represents the oscillations of the gapless mode. This additional factor
has the same form as the contribution of the phonons
to the superfluid He$_4$ wave function. Exactly as in the superfluid He$_4$,
the gapless mode factor gives a negligible small contribution to the ground
state energy but it is crucial to get the correct correlations. In fact,
MacDonald and Zhang \cite{mz} have recently calculated the collective
excitations for double-layer FQHE systems using the single mode approximation
on the Halperin ($m,m,m$) state \cite{halperin}, and find that it violates
the sum rules.

We calculate the quantum numbers of the quasiparticles. We find that the
charge of the quasiparticles for a given layer, is determined by the
filling fraction of the layer, and by the effective number of Landau levels
filled in that layer, at the mean field level of the semiclassical
approximation.
We compute the statistics of the quasiparticles. We show that it is well
define only in the case in which the coefficient of the Chern-Simons term
in the effective action for the fluctuations of the Chern-Simons gauge
fields, is not singular. In this case, the statistics of the quasiparticles
of a given layer is proportional to the number of fluxes attached to the
electrons in that layer. On the other hand, the relative statistics between
quasiparticles in different layers, is proportional to the relative number
of fluxes. In the case in which the statistics is not well defined, as
for instance for the ($m,m,m$) states, we find that the effect of the
gapless mode is to induce a long-range attraction that forces the
quasiparticles in different layers to move together, forming a
bound state.

We also develop a generalization of our theory in which the $SU(2)$ spin
symmetry is taken into account explicitly. We obtain the general form of
the $SU(2)$ hierarchies and show that, in general, they do not coincide
with those of the $U(1) \otimes U(1)$ theory.

The paper is organized as follows. In Section \ref{sec:CS} we discuss
the generalization of the fermionic Chern-Simons field theory to double-layer
systems, and derive the electromagnetic response functions. In Section
\ref{sec:colle} we calculate the spectrum of collective excitations for
the ($3,3,1$) state and for the ($m,m,m$) states. In Section \ref{sec:wave} we
show that the density response functions calculated within the gaussian
approximation saturate the $f$-sum rule.
We also derive the form of the absolute value squared of the ground state wave
function for these states, valid at long distances and in the thermodynamic
limit. In Section \ref{sec:Hall} we derive the Hall conductance and show that
it has the correct value already at the semiclassical level of our
 approximation.
In Section \ref{sec:stat} we discuss the quantum numbers of the
the quasiparticles. In Section \ref{sec:su2} we discuss the $SU(2)$
version of this theory. Finally, in Section \ref{sec:conc} we summarize
our results. Two appendices are devoted to the proof of
fermion-to-fermion Chern-Simons mapping in bilayers and to the estimate
of the contribution of the gaussian fluctuations of the ground state
energy of the $(1,1,1)$ state.

\section{Fermionic Chern-Simons theory for double layer FQHE systems}
\label{sec:CS}

In this section we describe the generalization of the Chern-Simons field
theory for the single-layer FQHE \cite{l1} to a double-layer two
dimensional electron system (2DES).

In the second quantized language, the action for a double-layer 2DES
in the presence of an external uniform magnetic field $B$ perpendicular to it
is given by
\begin{eqnarray}
{\cal S}&=&\int  d^3 z \; \sum _{\alpha}\; \left\{ \psi^*_{\alpha}(z)
[i D_0 +\mu_{\alpha}]
\psi_{\alpha}(z)-{1 \over2 M}|{\vec D}\psi_{\alpha}(z)|^2 \right\} \nonumber \\
&&-{1\over 2} \int  d^3z \int d^3z'\; \sum _{\alpha ,\beta}\
(|\psi_{\alpha}(z)|^2-{\bar\rho}_{\alpha})
   V_{\alpha \beta}(|{\vec z}-{\vec z}'|)
(|\psi_{\beta }(z')|^2-{\bar\rho}_{\beta})
\label{eq:ese1}
\end{eqnarray}
where the indices $\alpha =1,2$ and $\beta =1,2$ label the layers,
${\bar \rho}_{\alpha}$ is the average particle density in the layer
$\alpha$, $\psi(z)_{\alpha}$ is a second
quantized Fermi field, $\mu_{\alpha}$ is the chemical potential and $D_{\mu}$
is the
covariant derivative which couples the fermions to the external electromagnetic
field $A_{\mu}$. In what follows we
will assume that the pair potential has either the Coulomb form, {\it i.e.},
\begin{equation}
V(|\vec r|)_{\alpha\beta} =  {q^2 \over {\sqrt{{\vec r\;}^2 + {\vec d \;}^2 (1-
\delta_{\alpha\beta}))}}}
\label{eq:int}
\end{equation}
(with $d$ the interlayer separation), or that it represents a short range
interaction such that in momentum space it satisfies that
${V}_{\alpha \beta}(\vec Q){\vec Q}^2$ vanishes at zero momentum.
This includes the case of
ultralocal potentials ({\it i.e.}, with a range smaller or of the same order as
the cyclotron length $\ell$), in which case we can set ${\tilde V}(0) =
0$,
or short range potentials with a range longer than $\ell$ such as a
Yukawa interaction.

Following the same steps as in reference \cite{l1}, in the Appendix
\ref{sec:AA} we show that this system is equivalent to a
system of interacting electrons coupled to an additional statistical
vector potential
$a^{\mu}_{\alpha}$ ($\mu=0,1,2$) whose dynamics is governed by the Chern-Simons
action
\begin{equation}
{\cal S}_{\rm cs}= \sum _{\alpha\beta}\; {\kappa _{\alpha \beta}\over 2}
\int d^3x\;
\epsilon^{\mu \nu \lambda} {a}_{\mu}^{\alpha} \partial_{\nu}
{a}_{\lambda}^{\beta}
\label{eq:cs}
\end{equation}
provided that the CS coupling constant satisfies
\begin{eqnarray}
\kappa^{\alpha \beta}={1\over {2\pi (4 s_1 s_2 - n^2)}}
\left(
\begin{array}{cc}
2 s_2  & -n \\
-n & 2 s_1
\end{array}
\right)
\label{eq:Kalfabeta}
\end{eqnarray}
where $s_1$, $s_2$ and $n$ are arbitrary integers. In eq~(\ref{eq:cs})
$x_0,x_1 \;{\rm and}\; x_2$
represent the time and the space coordinates of the electrons respectively.
In the equivalent theory the covariant derivative is given by
\begin{equation}
D_{\mu}^{\alpha}=\partial_{\mu}+i{e\over c}A_{\mu}+i a_{\mu}^{\alpha}
\label{eq:dcov}
\end{equation}
and it couples the fermions to the statistical gauge fields
($a_{\mu}^{\alpha}$),
and to the external electromagnetic field ($A_{\mu}$). Notice that the
theory has now a $U(1) \otimes U(1)$ gauge invariance.

The Chern-Simons action implies a constraint for the particle density
$j_0^{\alpha}({\vec x})$ and the statistical flux
${\cal B}^{\alpha}={\epsilon }_{ij} {\partial}_{i} a_{j}^{\alpha}$, given by
\begin{equation}
j_0^{\alpha}({\vec x}) + \kappa _{\alpha \beta}  {\cal B}_{\beta}(\vec x) =0
\label{eq:cons}
\end{equation}
(from now on we assume that repeated indices are contracted).

This relation states that the electrons in plane ${\alpha}$ coupled to
statistical gauge fields with Chern-Simons coupling constant given by
eq~(\ref{eq:Kalfabeta}), see a statistical flux per particle of
$2\pi 2s_{\alpha}$ for the particles in their own plane , and
a statistical flux per particle of $2\pi n$ for the particles in the opposite
plane. (Notice that in units in which $e=c=\hbar=1$, the flux quantum is
equal to $2\pi$).
If the coefficient of the Chern-Simons
term is chosen with the above prescription, all the physical amplitudes
calculated in this theory are
identical to the amplitudes calculated in the standard theory, in which the
Chern-Simons field is absent. Of course, this is true provided that the
dynamics of the statistical gauge fields is fully taken into account exactly.

In this work we will take into account the dynamics of
the Chern-Simons gauge fields in a semiclassical
expansion, which is a sequence of well controlled approximations.
In practice, we will only consider the leading and next-to-leading order in the
semiclassical approximation.

Using the constraint enforced by the Chern-Simons action, the interaction term
of the action eq~(\ref{eq:ese1}) becomes
\begin{equation}
{\cal S}_{int}= - {1\over 2} \int  d^3z \int d^3z'\;
(\kappa ^{\alpha \delta} {\cal B}_{\delta}(z)-
{\bar\rho}_{\alpha})
V_{\alpha \beta}(|{\vec z}-{\vec z}'|)
(\kappa ^{\beta \gamma} {\cal B}_{\gamma}(z')-{\bar\rho}_{\beta})
\label{eq:eseint}
\end{equation}

The quantum partition function for this problem is, at zero temperature
\begin{equation}
{\cal Z}[A_{\mu}]=\int {\cal D} \psi^* {\cal D} \psi {\cal D}
a_{\mu}^{\alpha} \; \exp (i S(\psi^*,\psi,a_{\mu}^{\alpha},A_{\mu}))
\label{eq:pf}
\end{equation}
Since the action is quadratic in the fermions, they can be integrated out.
The effective action (${\cal S}_{\rm eff}$) is given by
(in units in which $e=c=\hbar=1$)
\begin{equation}
{\cal S}_{\rm eff}= -i \sum _{\alpha} {\rm tr} \; {\ln} \left\{
i D_0^{\alpha} +\mu_{\alpha}+
{1 \over2 M}({\vec D}^{\alpha})^2 \right\}
+ {\cal S}_{\rm cs} ( a_{\mu}^{\alpha} - {\tilde A}_{\mu}^{\alpha})
+ {\cal S}_{\rm eff}^{\rm int} ( a_{\mu}^{\alpha} - {\tilde A}_{\mu}^{\alpha})
\label{eq:ese2}
\end{equation}
where
\begin{eqnarray}
{\cal S}_{\rm eff}^{\rm int} ( a_{\mu}^{\alpha} - {\tilde A}_{\mu}^{\alpha}) =
-{1\over 2} \int  d^3z \int d^3z'&& \;(\kappa ^{\alpha \delta}
({\cal B}_{\delta}(z)-{\tilde B}_{\delta}(z))- {\bar\rho}_{\alpha}) \nonumber
\\
&& V_{\alpha \beta}(|{\vec z}-{\vec z}'|)
(\kappa ^{\beta \gamma} ({\cal B}_{\gamma}(z')-{\tilde B}_{\gamma}(z'))
-{\bar\rho}_{\beta})
\label{eq:esein2}
\end{eqnarray}
Here we have written the external electromagnetic field as
a sum of two terms, one representing the uniform magnetic
field $B$, and a small fluctuating term ${\tilde A}_{\mu}^{\alpha}$ whose
average
vanishes everywhere. The latter will be used to probe the electromagnetic
response of the system. Notice that we have used the invariance of the
measure ${\cal D} a_{\mu}^{\alpha}$ with respects to shifts, to move
${\tilde A}_{\mu}^{\alpha}$ out of the covariant derivatives and into the
Chern-Simons
and the interaction terms of the effective action (eq~(\ref{eq:ese2})).

\subsection{Mean field approximation: allowed fluid states}
\label{subsec:meanf}

The path integral $\cal Z$ can be approximated by expanding its degrees of
freedom in powers of the fluctuations, around stationary configurations of
${\cal S}_{eff}$.
This requirement yields the classical equations of motion.
\begin{eqnarray}
< j_0^{\alpha} (z)>_{F}&=& - {\kappa}_{\alpha\beta}[< {\cal B}^{\beta}(z)>-
{\tilde B}^{\beta}(z)>  \nonumber\\
< j_k^{\alpha} (z)>_{F}&=& - {\kappa}_{\alpha\beta} \epsilon_{kl}
[< {\cal E}^{\beta}_{l}(z)>-{\tilde E}^{\beta}_{l}(z)>] \nonumber \\
&& - \partial _{l}^{z} \epsilon_{lk} \int d^3 z' \; \kappa^{\alpha\epsilon}
V_{\epsilon\delta}(z,z') [-{\kappa}^{\delta\gamma}<{\cal B}^{\gamma}-
{\tilde B}^{\gamma}>(z')- {\bar \rho}^{\delta}]
\label{eq:sp}
\end{eqnarray}
Just as in the case of the single layer problem, these equations have many
possible
solutions, {\it i.e.}, fluid states, Wigner crystals and non-uniform states
with vortex-like configurations. We will only consider
solutions with uniform particle density, {\it i.e.}, the liquid phase solution.
This is the {\it average field approximation} (AFA), which can be
regarded as a mean field approximation. At the mean
field level the electrons in the layer $\alpha$ see a total flux
$B_{\rm eff}^{\alpha}$, equal to the
external magnetic flux partially screened by the average Chern-Simons flux,
{\it i.e.} $B_{\rm eff}^{\alpha}=B+\langle {\cal B}^{\alpha}\rangle
=B-({\kappa ^{-1})^{\alpha\beta} {\bar \rho}_{\beta}}$.
It is easy to see that the uniform saddle-point state has a gap only if the
effective field $B_{\rm eff}^{\alpha}$ is such that the fermions in layer
$\alpha$ fill exactly an integer number $p_{\alpha}$ of the effective
Landau levels, {\it i.e.}, those defined by $B_{\rm eff}^{\alpha}$.
In other words, the AFA to this theory yields a state with an energy gap if
the filling fractions of each layer satisfy
\begin{eqnarray}
\nu _{1} &=& {1 \over {\pm {1\over {p_{1}}}+ 2s_{1} + n {N_{2}\over N_{1}}}}
\nonumber \\
\nu _{2} &=& {1 \over {\pm {1\over {p_{2}}}+ 2s_{2} + n {N_{1}\over N_{2}}}}
\label{eq:fill}
\end{eqnarray}
where $N_{1}$ and $N_{2}$ are the number of particles in layers $1$ and $2$
respectively. The sign in front of $p_{1}$ and $p_{2}$ indicates if the
effective field is parallel or antiparallel to the external magnetic field.

Using the fact that $\nu = \nu_{1} + \nu_{2}$, and that the number of flux
quanta enclosed by each plane is the same, the filling fractions can be
written as follows
\begin{eqnarray}
\nu _{1} &=& {{n-(\pm {1\over p_{2}} + 2 s_{2})} \over
              { n^{2} - (\pm {1\over p_{1}} + 2 s_{1})(\pm {1\over p_{2}}
+ 2 s_{2})}}
\nonumber \\
\nu _{2} &=& {{n-(\pm {1\over p_{1}} + 2 s_{1})} \over
              { n^{2} - (\pm {1\over p_{1}} + 2 s_{1})(\pm {1\over p_{2}}
+ 2 s_{2})}}
\label{eq:fillfrac}
\end{eqnarray}
where $p_{1}$, $p_{2}$, $s_{1}$, $s_{2}$ and $n$ are arbitrary integers.
For the special case in which the two layers have the same occupancy,
$N_1=N_2$ and $\nu_1=\nu_2={\frac{\nu}{2}}$, the allowed fractions are
\begin{equation}
\nu(p,n,s)={\frac{2p}{(n+2s)p+1}}
\label{eq:simplerfraction}
\end{equation}
where $p$ is an arbitrary (positive or negative) integer.

The effective magnetic field can be written in terms of the external
magnetic field as
\begin{equation}
B_{\rm eff}^{\alpha}= B { \nu^{\alpha} \over  p^{\alpha}}
\label{eq:beff}
\end{equation}
For general values of these integers, the states whose filling fractions
are given by eq~ (\ref{eq:fillfrac}) have a
gap at the mean field level of this approximation. This ensures that the
perturbative expansion is meaningful. In other words, there is a small
parameter, which is essentially the inverse of the mean field gap, for
this perturbative expansion to be possible.
If there is no gap for the excitations of the mean field ground state,
the perturbative expansion breaks down.
The breakdown is signalled by infrared divergencies at low temperatures.
Such is the case for the compressible or ``Fermi Liquid" states. We will
see now that both types of states can occur for a given filling fraction
(but not for all filling fractions!).

It is clear from eq.~(\ref{eq:fillfrac}) that in the bilayer systems, there are
many possible choices of the numbers $p_{1}$, $p_{2}$, $s_{1}$, $s_{2}$ and
$n$, {\it i.~e.~,} many different states, which have the same filling
fractions.
As a result, the phase diagram for bilayers is much richer than for
spin polarized electrons in a single layer.
Experiments on spin polarized 2DEG bilayers\cite{e2,unmed} at
filling fractions $\nu=1,1/2$ have
shown a non-trivial phase diagram with at least two phases: a
compressible phase in the regime in which the 2DEG's are well separated
and an incompressible phase when they are closer by.  However, as it was
emphasized by Wen and Zee\cite{wen}, the
incompressible states at these two filling fractions actually have quite
different properties. We will see now that the phase diagram can be
quite complex.

Let us consider some cases of special interest, in particular with
$N_1=N_2$. For example, at level one of the hierarchy, we choose
$p_{1}=p_{2}=1$, $\; 2s_{1}+1 = m_{1}$, $\; 2s_{2}+1
= m_{2}$, and $n=m$, and we obtain the so called $(m_1, m_2, m)$ states
\cite{wen,halperin,wilczek} whose filling fractions
are $\nu = {m_{1} + m_{2} - 2n \over {m_{1} m_{2} - n^2}}$. In
particular,
for $m_{1}=m_{2}=3$ and $m=1$, this is the $(3,3,1)$ state, whose filling
fraction is $\nu = {1\over 2}$.
For $p_{1}=p_{2}=1$, and $2s_{1}+1=2s_{1}+1=n=m$, we obtain the $(m, m, m)$
states \cite{wen,ezawa} whose filling fractions are $\nu = {1 \over m}$.

We can also consider the limit $p_{1}=p_{2}\equiv p \rightarrow \infty$.
In this case, if $s_{1}=s_{2}=s$, the state with filling fraction $\nu={1
\over m}$ will be obtained for all the values of $n$ and $s$ such that
$n+2s= 2m$.
Since $p \rightarrow \infty$, the effective field vanishes, and we find the
analogous of the compressible states for the single layer problem discussed
by Halperin et al \cite{hlr}.
All of these states are degenerate with the
$\nu ={1\over m}$ states mentioned above.

Clearly, it is always possible to construct a large family of, in
principle, distinct
states which have the same filling fraction. For instance, we can also
choose $p_{1}=p_{2}=1$, $n=-m$ and $2s_{1}+1=2s_{2}+1=3m$. For the case
$N_1=N_2$, it is easy to see that all the states with the same value of
$n+2s$ have the same filling fraction (at fixed $p_1=p_2=p$). For
$N_1 \not= N_2$, similar (but more complicated) families can be found.
Let $(N_1, N_2)$ denote the largest common divisor ({\it l.~c.~d.~}) of
$N_1$ and $N_2$ and let $\ell$ and $k$ be two relatively prime integers
({\it i.~e.~} $(\ell,k)=1$) such that $N_1=\ell (N_1, N_2)$ and  $N_2=k
(N_1, N_2) $.
By direct inspection of the expressions for
the filling fractions of eq.~(\ref{eq:fill}), we find that the
following transformations leave the filling fractions $\nu_1$ and
$\nu_2$ invariant:
\begin{eqnarray}
n    & \to & n   - 2rk \ell \nonumber \\
s_1  & \to & s_1 +  rk^2      \nonumber \\
s_2  & \to & s_2 +  r  \ell^2
\label{eq:degeneracy}
\end{eqnarray}
where $r$ is an arbitrary integer.

In simpler terms,
new states are generated by transferring {\it intra}-plane flux-particle
attachment, determined by $s_1$ and $s_2$, to {\it inter}-plane
flux attachment, determined by $n$. Since the inter-plane
attached fluxes are negative in these states, the particles of one
plane see the particles of the other plane as if they were holes. Thus,
in these states, there is an effective attractive force between
particles
on different planes. Hence, in these states the wave functions, instead
of having zeros of higher order (which represent repulsion), should have
a larger weight when particles from different
planes are closer to each other (in the sense of their coordinates
projected on the $xy$ plane). Asymptotically, these wave functions
appear to have ``poles" in the interlayer coordinates. Below we find
that this is indeed the case.

Which one
of all of these states is realized for a given system at a fixed filling
fraction should depend on the
inter and intra layer interactions. At the mean field level we find that
these states are all degenerate.
In appendix ~\ref{sec:BB} we give an estimate of the
contribution of the gaussian
corrections to the ground state energy. Our results show that
this degeneracy is lifted and that, of all the incompressible states at
filling fraction $1/m$, the simpler $(m,m,m)$
states have lower energy. Nevertheless, the situation should be in
principle more complex. For instance, one can imagine an interlayer
interaction which is very weak at short in-plane distances but stronger
at intermediate separations (``hollow core"). In such cases, the
off-plane attachment mechanism might prevail. In addition, for truly
large interlayer separations, the system should prefer to have (for
$\nu=1/m$) a state with a ``Fermi Liquid" state on each layer. A
detailed calculation of
the contribution of fluctuations to the ground state energy is necessary
to establish which one of these states actually occurs for a given form
of the interaction and interlayer separation. We will return to this
problem elsewhere.

As it stands, this
theory has no a priori way of limiting the number of possible states.
Clearly, many of the states in these families must be redundant since,
for a fixed number of particles and fluxes, the dimension of the Hilbert
space is fixed. Hence, either by virtue of
a symmetry the states in these families fall
into equivalency classes, or there should be a natural way to limit the
number of states. Fluctuations must also play an important role here.
Already at the gaussian level the fluctuations do lift these
degeneracies.  However, at the present time, it is unclear how will the
fluctuations manage to reduce the size of the Hilbert space.

A more
natural resolution of these issues would be that a physical
interpretation of these new families of states could be found in terms
of the physical degrees of freedom of the electrons. It is natural to
speculate that the degeneracy found here may become natural
in the spin picture of the bilayer system. Thus, if we
think the electrons in the upper layer as having spin up while those in
the lower layer have spin down, the states can now be classified in
terms of the total number of electrons (charge), and of the difference
of occupancy, which can be viewed as the $z$-projection $S_z$ of the
``spin". All of the states that we have discussed here have the same
occupancy in both planes and, hence, they have $S_z=0$. In the
limit of small separations, the system should have an effective $SU(2)$
symmetry even if the electrons are fully polarized. In that limit, there
is an additional operator which commutes with the Hamiltonian, ${\vec
S}^2$, the ``total spin". Even though for an arbitrary interlayer
interaction the $SU(2)$ symmetry is broken down to $U(1)$, the $SU(2)$
states can still be used to enumerate the states (even though they are
no longer good quantum numbers). Thus, it is natural to think that the
multiple solutions may be linear combinations of the $S_z=0$ states with
different total ``spin" $S\leq 2N$. For the counting of states to be the
same, it is necessary that the families of states found above
either terminate (and be finite) or become split into equivalence
classes. In this system, the $SU(2)$-invariant limit is achieved either
when the two layers physically coincide or if the electrons are
non-interacting. However,
the two-component Chern-Simons gauge theory that we use here does not
exhibit the $SU(2)$ symmetry explicitly even in the limit of zero
interlayer separation. In this theory, the $SU(2)$ symmetry is a
dynamical symmetry. In other terms, given that this is an exact
transcription of the two-component 2DEG, the exact states should form
$SU(2)$ multiplets. However, while this is a property of the exact
theory, it is by no means guaranteed that a semiclassical approach would
be able to recover this exact property. Such dynamical mechanisms
are known to work in the abelian bosonization of $SU(2)$ invariant
$1+1$-dimensional Fermi systems\cite{luther}.

Finally, since the bilayer system can also be used to describe
spin unpolarized electrons, the allowed bilayer states should include
partially polarized and unpolarized (singlet) spin states. One problem
in the way
of making this connection  is the fact that this formalism breaks down
if the Chern-Simons coupling matrix is singular. This happens for $n=\pm
2s$. Thus, the spin singlet state $(3,3,2)$, which has
filling fraction $\nu=2/5$, cannot be described within this abelian
Chern-Simons approach. In the section \ref{sec:su2} we present a
generalization
of the theory of Balatsky and Fradkin\cite{su2} which describes the
$SU(2)$ symmetric cases.

\subsection{Semiclassical approximation (RPA)}
\label{subsec:sem}

Now we consider the gaussian (or semiclassical) fluctuations
of the statistical vector potential ${\tilde a}_{\mu}^{\alpha}$ around the
mean-field state. The gaussian corrections must alter the qualitative
properties of the state described by the AFA, which violates
explicitly Galilean invariance (more generally, {\it magnetic invariance})
which, for translationally invariant
systems, must remain unbroken and unchanged. Thus the center of mass of
the system must execute a cyclotron-like motion at, exactly, the
cyclotron
frequency of non interacting electrons in the full external magnetic
field, as demanded by Kohn's theorem\cite{kohn}. A na{\"\i}ve
application of the AFA would suggest that the cyclotron frequency is
renormalized downwards since the effective field seen by the composite
fermions is smaller than the external field $B$. Hence, the {\it
magnetic algebra} may appear to have changed. In the same way as it happens
for the single-layer systems \cite{l3}, here the
gaussian fluctuations yield the correct cyclotron frequency and,  thus,
restore the correct magnetic algebra.

At the gaussian level, the effective action for ${\tilde a}_{\mu}^{\alpha}$
is
\begin{eqnarray}
S_{\rm eff}({\tilde a}_{\mu}^{\alpha},{\tilde A}_{\mu}^{\alpha})&=&
      {1\over 2}  \int d^{3}x \int d^{3}y \; {\tilde a}_{\mu}^{\alpha}(x)\;
                  \Pi^{\mu \nu}_{\alpha\beta}(x,y) \;
                {\tilde a}_{\nu}^{\beta}(y)
\nonumber \\
&&-{1\over 2} \int  d^3x \int d^3y \;({\cal B}_{\delta}(x)-
{\tilde B}_{\delta}(y)) \kappa ^{\alpha \delta}
V_{\alpha \beta}(|{\vec x}-{\vec y}|)
\kappa ^{\beta \gamma} ({\cal B}_{\gamma}(y)-{\tilde B}_{\gamma}(y))
\nonumber \\
&&+ {\kappa _{\alpha \beta}\over 2} \int d^3x\;\epsilon^{\mu \nu \lambda}
({\tilde a}_{\mu}^{\alpha}-{\tilde A}_{\mu}^{\alpha}) \partial_{\nu}
({\tilde a}_{\lambda}^{\beta}-{\tilde A}_{\lambda}^{\beta})
\label{eq:gauss}
\end{eqnarray}
The tensor $\Pi^{\mu \nu}_{\alpha\beta}(x,y)=\Pi^{\mu \nu}_{\alpha}(x,y)
\delta_{\alpha\beta}$, where $\Pi^{\mu \nu}_{\alpha}(x,y)$ is the polarization
tensor of the
equivalent fermion problem at the mean field level. It is obtained by
expanding the fermion determinant up to quadratic order in the
statistical gauge field. This tensor was calculated in reference
\cite{l1}. The subindex $\alpha$ indicates that the effective field which
appears in the expressions for  $\Pi^{\mu \nu}$ in reference \cite{l1} is
$B_{\rm eff}= B_{\rm eff}^{\alpha}$.

After integrating out the gaussian fluctuations of the statistical gauge
fields ${\tilde a}_{\mu}^{\alpha}$, we obtain the effective action for the
electromagnetic fluctuations (${\tilde A}_{\mu}^{\alpha}$),
$S_{\rm eff}^{\rm em}({\tilde A}_{\mu}^{\alpha})$
\begin{equation}
{\cal S}_{\rm eff}^{\rm em} ({\tilde A}_{\mu}^{\alpha}) = {1\over 2}
                  \int d^{3}x \int d^{3}y {\tilde A}_{\mu}^{\alpha}(x)
                  K^{\mu \nu}_{\alpha\beta}(x,y)
                   {\tilde A}_{\nu}^{\beta} (y)
\label{eq:ems}
\end{equation}
Here $K^{\mu\nu}_{\alpha\beta}$ is the electromagnetic polarization tensor.
It measures the
linear response of the system to a weak electromagnetic perturbation.

We will
use this effective action to calculate the full electromagnetic response
functions at the gaussian level. Since this calculation is based on a
one loop effective action for the fermions ({\it i.e.} a sum of fermion
bubble diagrams), this approximation amounts
to a random phase correction to the average field approximation.

The components of the electromagnetic polarization tensor can be written in
momentum space as follows
\begin{eqnarray}
K_{00}^{\alpha\beta} &=& {\vec Q}^2  K_{0}^{\alpha\beta}(\omega, {\vec Q})
\nonumber \\
K_{0j}^{\alpha\beta} &=& {\omega} Q_{j} K_{0}^{\alpha\beta}(\omega, {\vec Q})
           + i {\epsilon _{jk}}Q_{k} K_{1}^{\alpha\beta}(\omega, {\vec Q})
\nonumber  \\
K_{j0}^{\alpha\beta} &=& {\omega} Q_{j} K_{0}^{\alpha\beta}(\omega, {\vec Q})
           - i {\epsilon _{jk}} Q_{k} K_{1}^{\alpha\beta}(\omega, {\vec Q})
\nonumber \\
K_{ij}^{\alpha\beta} &=& {\omega}^2 {\delta _{ij}} K_{0}^{\alpha\beta}(\omega,
  {\vec Q}) - i {\epsilon _{ij}} {\omega} K_{1}^{\alpha\beta}(\omega, {\vec Q})
+({\vec Q}^2 {\delta _{ij}}- {Q_i}{Q_j})K_{2}^{\alpha\beta}(\omega, {\vec Q})
\label{eq:respfnt}
\end{eqnarray}
where $K_{i}^{\alpha\beta}(\omega, {\vec Q})$ ($i=0,1,2$) can be written as
a power series expansion in powers of ${\vec Q}^2 \over {B_{\rm eff}^{\alpha}}$
and have poles for the values of $\omega$ that coincide with the collective
modes of the system.

The electromagnetic response functions determined by $K_{\mu\nu}^{\alpha\beta}$
 have the following properties:

{i)} As in the single-layer case, the collective excitations of this systems
are
determined by the poles of the density correlation function,
$K_{00}^{\alpha\beta}(\omega, {\vec Q})$.

{ii)}The leading order term in ${\vec Q}^2$ of the $K_{00}^{\alpha\beta}$
component of the polarization tensor saturates the $f$-sum rules. These sum
rules correspond to the conservation of the particle number in each layer
separately.
This result is essential in order to show that the absolute value squared of
the
ground state wave functions of all the (incompressible) liquid states have
the form described in the introduction at very long distances and
in the thermodynamic limit.

{iii)} The gaussian fluctuations of the statistical gauge field are
responsible
for the FQHE. In particular, the gaussian corrections yield the exact
value for the Hall conductance.

In the next sections we will discuss these properties in detail.

\section{Spectrum of collective excitations}
\label{sec:colle}

In this section we derive the spectrum of collective excitations for two
different states, the $(m,m,n)$ and the $(m,m,m)$ states.
We use the same approach as we did for
the single-layer case \cite{l3}, i.e., we study the poles of the
density correlation function to determine the collective modes of the system.

\subsection{Collective excitations for $(m,m,n)$ states}
\label{subsec:unmedio}

For simplicity, we have studied the collective modes for the state
$(3,3,1)$. All the other states can be studied by straightforward
application of the same methods.

In this case the total filling fraction is $\nu = {1\over 2}$,
being ${\nu}_{1} = {\nu}_{2} = {1\over 4}$ . The effective
cyclotron frequencies and magnetic fields are
${\omega}_{\rm eff}^{1}={\omega}_{\rm eff}^{2}=
{\bar \omega}= {{\omega}_{c} \over 4}$ and ${B}_{\rm eff}^{1}={B}_{\rm eff}^{2}
={\bar B}= {B \over 4}$.

We find that there is a family of collective modes whose zero-momentum
gap is $k {\bar \omega}$, where $k$ is an integer number different from
$1$. At mean field level, there are two modes for each
integer multiple of ${\bar \omega}$. After including the gaussian fluctuations
we find
that there are no modes with a  zero momentum gap at ${\bar \omega}$.
One of them has been ``pushed up" to the cyclotron frequency and the
other up to $2 {\bar \omega}$ (at ${\vec Q}=0$). Therefore, at these
multiples of  $ {\bar \omega}$ there are three degenerated modes
for ${\vec Q}=0$.
For ${\vec Q} \not=0$, the degeneracy is lifted and these three modes have
different dispersion curves.

At $2 {\bar \omega}$ we find that there are two modes with residue
${\vec Q}^2$, and one with residue ${\vec Q}^4$. The former are
\begin{equation}
\omega_{\pm} ({\vec Q})= \Big [ (2 {\bar \omega})^2 +
                         ({{\vec Q}^2\over 2 {B_{\rm eff}}})^{1\over 2}\;
                         {\bar \omega}^2 \; {\alpha }_{\pm}
             \Big ] ^{1\over2}
\label{eq:uno}
\end{equation}
where
\begin{equation}
{\alpha}_{\pm} ={3M\over 2\pi} ({v}_{11}-{v}_{12}) \pm
            {\sqrt {({3M\over 2\pi})^{2}({v}_{11}-{v}_{12})^2 +16}}
\label{eq:unosub}
\end{equation}
Here $v_{\alpha\beta}$ are the zeroth order coefficients of the
Fourier transform of the interparticle pair potential for short range
interactions. For Coulomb interactions $ v_{11}= {q^2 \over \epsilon}$ and
$ v_{12}= {q^2 \over \epsilon}{e^{-|{\vec Q}|d}}\approx {q^2 \over \epsilon}$
if $|{\vec Q|}d \ll 1$, therefore ${\alpha }_{\pm}= \pm 4$ in this limit.

The residues in $K_{00}^{\alpha\beta}$ corresponding to these poles are
\begin{equation}
Res(K_{00}^{\alpha\beta},\omega _{\pm}({\vec Q})) = -{\vec Q}^2 \;
                                   {\omega}_{c} {\nu \over 8\pi}\;
            \left( \begin{array}{cc} 1& \;  -1\\ -1& \; 1
                   \end{array} \; \right)
                     ( 1+ {16\over {\alpha}_{\pm}^{2}} )^{-1}
\label{eq:resuno}
\end{equation}
It is clear from the form of these residues, that
these excitations are {\it out of phase} modes, because they only couple
the "out of phase" density (i.e., they couple ${\rho }^{-}$ with itself).

The other mode at $2 {\bar \omega}$ is
\begin{equation}
\omega_{0} ({\vec Q})= \Big [ (2 {\bar \omega})^2 -
                         6 ({{\vec Q}^2\over 2 {B_{\rm eff}}})\;
                         {\bar \omega}^2
             \Big ] ^{1\over2}
\label{eq:dos}
\end{equation}
and its residue is proportional to ${\vec Q}^4 \left( \begin{array}{cc} 1& \;
  1\\ 1& \; 1   \end{array} \; \right) $. Thus, this is an {\it in phase}
mode since it only couples ${\rho }^{+}$ with itself.

The two modes with zero momentum frequency $3 {\bar \omega}$ are given by
\begin{equation}
\omega ({\vec Q})= \Big [ {3 \bar \omega}^2 +6
                 ({{\vec Q}^2\over 2 {B_{\rm eff}}})^{2}\;
                         {\bar \omega}^2 \;
            \Big ] ^{1\over2}
\label{eq:cuatro}
\end{equation}
with residue proportional to ${\vec Q}^6\left(
\begin{array}{cc} 1& \;  -1\\ -1& \; 1   \end{array} \; \right)$ (this is an
{\it out of phase} mode); and
\begin{equation}
\omega ({\vec Q})= \Big [ {3 \bar \omega}^2 -18
                 ({{\vec Q}^2\over 2 {B_{\rm eff}}})^{2}\;
                         {\bar \omega}^2 \;
            \Big ] ^{1\over2}
\label{eq:cinco}
\end{equation}
with residue proportional to ${\vec Q}^6\left(
\begin{array}{cc} 1& \;  1\\ 1& \; 1   \end{array} \; \right)$ (this is an
{\it in phase} mode).

There are three modes whose zero momentum frequency is the cyclotron frequency.
\begin{equation}
\omega ({\vec Q})= \Big [ { \omega}_{c}^2 +
                         ( 2+ {M( v_{11} + v_{12}) \over \pi})
                        3  ({{\vec Q}^2\over 2 {B_{\rm eff}}})\;
                         {\bar \omega}^2   \Big ] ^{1\over2}
\label{eq:tres}
\end{equation}
with residue
\begin{equation}
Res(K_{00},\omega ({\vec Q})) = -{\vec Q}^2 \;
                                   { \omega}_{c} {\nu \over 8\pi}\;
            \left( \begin{array}{cc} 1& \;  1\\ 1& \; 1
                   \end{array}\; \right) ,
\label{eq:restres}
\end{equation}
\begin{equation}
\omega ({\vec Q})= \Big [ { \omega}_{c}^2 -
                        16 ({{\vec Q}^2\over 2 {B_{\rm eff}}})^{2}\;
                         {\bar \omega}^2 \;
                ( 1+ {3M( v_{11} + v_{12}) \over 2\pi})^{-1}
                                  \Big ] ^{1\over2}
\label{eq:seis}
\end{equation}
with residue proportional to ${\vec Q}^4 \left(
\begin{array}{cc} 1& \;  1\\ 1& \; 1   \end{array} \; \right)$
\begin{equation}
\omega ({\vec Q})= \Big [ {\omega}_{c}^2 + 2
                 ({{\vec Q}^2\over 2 {B_{\rm eff}}})^{3}\;
                         {\bar \omega}^2 \;
            \Big ] ^{1\over2}
\label{eq:siete}
\end{equation}
with residue proportional to ${\vec Q}^8 \left(
\begin{array}{cc} 1& \;  -1\\ -1& \; 1   \end{array} \; \right)$.
The first two modes at ${\omega}_{c}$ are {\it in phase} and the last one is
{\it out of phase}.

In summary, we find a family of collective modes with dispersion
relations
whose zero-momentum gap is $k{\bar \omega}$, where $k$ is an
integer number
different from $1$. When $k=4$, {\it i.e.} the zero-momentum gap is the
cyclotron frequency, there is a splitting in the dispersion relation for
finite wavevector. However, only one of these three modes has residue
proportional to ${\vec Q}^2$ in the density correlation function. One expects
that the other modes will become damped
due to non-quadratic interactions among the collective modes. On the other
hand, the collective mode with lowest energy which has $k=2$, is stable
(at least for reasonably small wavevectors).

The validity of the spectrum described in this section is limited by the
fact that we have not considered the physics at arbitrary wavevectors,
and the (expected) effects of non-gaussian corrections.
At the gaussian (RPA) level and for small momentum, we found a family of
collective modes which are infinitely long lived, ({\it i.e.}, the response
functions have delta-function sharp poles at their location).
These modes represent charge-neutral bound states.
For $\vec Q$ sufficiently large, the energy of the collective mode can
become equal to the energy  necessary to create the lowest available
two-particle state: a quasiparticle-quasihole pair. At this point, the
collective mode should become damped.
Non-gaussian corrections to the RPA are also expected to give a finite
width to (presumably) all the collective modes but the lowest one.
Since the modes with zero momentum gap at
$k{\bar \omega}$, $k \geq 3$, are not the collective modes with lowest
energy, it is possible that at finite wavevectors they may also decay into
the collective mode with lowest energy ( the mode with
$k=2$, which has a gap at ${\bar \omega}$).

\subsection{Collective excitations for $(m,m,m)$ states}
\label{subsec:ememe}

Here we present the spectrum of collective excitations for the so
called $(m,m,m)$ states.
In this case the total filling fraction is $\nu = {1\over m}$,
being ${\nu}_{1} = {\nu}_{2} = {1\over 2m}$ . The effective
cyclotron frequencies and magnetic fields are
$\; {\omega}_{\rm eff}^{1}={\omega}_{\rm eff}^{2} \equiv
{\bar \omega}= {{\omega}_{c} \over 2m}$ and
${B}_{\rm eff}^{1}={B}_{\rm eff}^{2} \equiv {B _{\rm eff}} = {B \over 2m}$.

We find again a family of collective modes whose zero-momentum
gap is $k {\bar \omega}$, where $k$ is an integer number different from
$1$. At mean field level, there are two modes for each
integer multiple of ${\bar \omega}$. After including the gaussian fluctuations
we find
that there are no modes with a  zero momentum gap at ${\bar \omega}$.
One of them has been ``pushed up" to the cyclotron frequency. Therefore, at
$ {\omega}_{c}$ there are three degenerate modes for ${\vec Q}=0$.
For ${\vec Q} \not=0$, the degeneracy is lifted and these three modes have
different dispersion curves.
The other mode at ${\bar \omega}$
has been ``pulled down"  to zero frequency at ${\vec Q}=0$, i.e., it has become
a gapless mode.

We will distinguish between the cases $m=1$ and $m \not= 1$.

\noindent a) Case $m=1$

The gapless mode is
\begin{equation}
\omega ({\vec Q})= {v_{s}} |{\vec Q}|
\label{eq:veinte}
\end{equation}
where
\begin{equation}
{v_{s}}^{2}=\Big [ 1+ {M\over 2\pi}(v_{11} - v_{12})\Big ]
{ {\omega}_{c} \nu \over 2M}
\label{eq:vel}
\end{equation}
where $v_{\alpha\beta}$ are the zeroth order coefficient of the
Fourier transform of the interparticle pair potential for short range
interactions.
For Coulomb interactions $ v_{11}({\vec Q})= {q^2 \over \epsilon}$ and
$ v_{12}({\vec Q}) = {q^2 \over \epsilon}{e^{-|{\vec Q}|d}}\approx {q^2 \over
\epsilon}$
if $|{\vec Q}|d \ll 1$. Therefore, $(v_{11} - v_{12})=0$ for Coulomb
interactions (in the limit $|{\vec Q}|d \ll 1$).

The residue in $K_{00}^{\alpha\beta}$ corresponding to this pole is
\begin{equation}
Res(K_{00},\omega ({\vec Q})) = -{\vec Q}^2 \;
                                   {\omega}_{c} {\nu \over 8\pi}\;
            \left( \begin{array}{cc} \;1&   -1\\ -1& \; 1
                   \end{array} \; \right)
\label{eq:resveinte}
\end{equation}
Therefore, this is an {\it out of phase} mode.

At ${\omega }_{c}=2 {\bar \omega}$ we find that there are two
({\it in phase}) modes
with residue ${\vec Q}^2$
\begin{equation}
\omega_{\pm} ({\vec Q})= \Big [  {\omega}_{c}^2 +
                         ({{\vec Q}^2\over 2 {B_{\rm eff}}})^{1\over 2}\;
                         {\bar \omega}^2 \; {\alpha }_{\pm}
             \Big ] ^{1\over2}
\label{eq:veintiuno}
\end{equation}
where
\begin{equation}
{\alpha}_{\pm} = {M\over 2\pi}({v}_{11}+{v}_{12}) \pm
            {\sqrt {({M\over 2\pi})^{2}({v}_{11}+{v}_{12})^2 +16}}
\label{eq:veintiunosub}
\end{equation}
 For Coulomb interactions $ v_{11} +v_{12}=  2 {q^2 \over \epsilon}$ if
$|{\vec Q}|d \ll 1$, therefore this term is higher order in $\vec Q$ and it
should be neglected, i.e., ${\alpha }_{\pm}=4$.

The residues in $K_{00}^{\alpha\beta}$ corresponding to these poles are
\begin{equation}
Res(K_{00},\omega _{\pm}({\vec Q})) = -{\vec Q}^2 \;
                                   {\omega}_{c} {\nu \over 8\pi}\;
            \left( \begin{array}{cc} 1& \;  1\\ 1& \; 1
                   \end{array} \; \right)
                     ( 1+ {16\over {\alpha}_{\pm}^{2}} )^{-1}
\label{eq:resveintiuno}
\end{equation}
therefore these are {\it in phase} modes.
The other mode at $\omega _{c}$ is
\begin{equation}
\omega_{0} ({\vec Q})= \Big [ {\omega}_{c}^2 - 2
                       ({{\vec Q}^2\over 2 {B_{\rm eff}}})\;
                         {\bar \omega}^2      \Big ] ^{1\over2}
\label{eq:veintidos}
\end{equation}
and its residue is proportional to ${\vec Q}^4
\left( \begin{array}{cc} \; 1&  - 1\\ -1& \; 1
                   \end{array} \; \right)$ ({\it out of phase} mode).

The modes with zero momentum frequency $\omega = k {\bar \omega}$ for
$k \geq 3$ coincide with the expressions given below for the case $m \not= 1$,
in eq~(\ref{eq:treinteycinco}) and eq~(\ref{eq:treinteyseis}).

\noindent b) Case $m \not= 1$

The gapless mode is an {\it out of phase} mode with the same form as for $m=1$
(eq~(\ref{eq:veinte}))
and with the same residue (eq~(\ref{eq:resveinte})).

At ${\omega }_{c}=2m {\bar \omega}$ we find that there is one {\it in phase}
mode with residue ${\vec Q}^2$
\begin{equation}
\omega ({\vec Q})= \Big [  {\omega}_{c}^2 +
              \left( {{2m-1}\over {m-1}}+ {M\over 2\pi}(v_{11} + v_{12})\right)
                   2m  ({{\vec Q}^2\over 2 {B_{\rm eff}}})\;
                         {\bar \omega}^2               \Big ] ^{1\over2}
\label{eq:treintayuno}
\end{equation}

The residue in $K_{00}^{\alpha\beta}$ corresponding to this pole is
\begin{equation}
Res(K_{00},\omega ({\vec Q})) = -{\vec Q}^2 \;
                                   {\omega}_{c} {\nu \over 8\pi}\;
            \left( \begin{array}{cc} 1& \;  1\\ 1& \; 1
                   \end{array} \; \right)
\label{eq:restreintayuno}
\end{equation}

The other modes at $\omega _{c}$ are one {\it in phase} mode
\begin{equation}
\omega ({\vec Q})= \Big [ {\omega}_{c}^2 -
                      {{4 m^{2} (2m-1)^{2} (2m-2) }\over
     {{(2m-2)!}\left( (2m-1) + (m-1){M\over 2\pi}(v_{11} + v_{12})\right) }}
                       ({{\vec Q}^2\over 2 {B_{\rm eff}}})^{2m-2}\;
                         {\bar \omega}^2  \Big ] ^{1\over2}
\label{eq:treinteytres}
\end{equation}
with residue proportional to ${\vec Q}^{4(m-1)}$, and one {\it out of phase}
mode
\begin{equation}
\omega ({\vec Q})= \Big [ {\omega}_{c}^2 -  {2 \over {(2m-2)!}}
                        ({{\vec Q}^2\over 2 {B_{\rm eff}}})^{2m-1}\;
                         {\bar \omega}^2 \Big ] ^{1\over2}
\label{eq:treinteycuatro}
\end{equation}
with residue proportional to ${\vec Q}^{4m}$.

The two modes with zero momentum frequency $k {\bar \omega}$ ($k \not= 1,2m$)
are given by
\begin{equation}
\omega ({\vec Q})= \Big [ {k \bar \omega}^2 -
                           {2 \over {(k-2)!}}
                        ({{\vec Q}^2\over 2 {B_{\rm eff}}})^{k-1}\;
                         {\bar \omega}^2 \Big ] ^{1\over2}
\label{eq:treinteycinco}
\end{equation}
and
\begin{equation}
\omega ({\vec Q})= \Big [ {k \bar \omega}^2 -
                           {2 k (2m-1) \over {(k-2)! (2m-k)}}
                        ({{\vec Q}^2\over 2 {B_{\rm eff}}})^{k-1}\;
                         {\bar \omega}^2 \Big ] ^{1\over2}
\label{eq:treinteyseis}
\end{equation}
Their residues are proportional to
${\vec Q}^{2k}\left( \begin{array}{cc} \; 1&   -1\\ -1& \; 1
                   \end{array} \; \right)$,
and to ${\vec Q}^{2k} \left( \begin{array}{cc} 1& \;  1\\ 1& \; 1
                   \end{array} \; \right)$ respectively.

In summary,  for the $(m,m,m)$ states, in addition to a family of collective
modes with dispersion relations whose zero-momentum gap is $k{\bar \omega}$
($k\not= 1$), we find that there is a gapless mode.
The gapless mode is related to the relative fluctuations of the
electronic density in each layer, i.e., to the fluctuations of
${\bar \rho}^1 - {\bar \rho}^2$.

All the considerations about the validity of this spectrum of collective
excitations beyond the semiclassical approximation that we discussed in the
previous section are of course valid in this case.

\section{Ground state wave function}
\label{sec:wave}

We need to show first that the long wavelength form of $K_{00}^{\alpha\beta}$,
found at this
semiclassical level, saturates the $f$-sum rule. This result implies
that the non-gaussian corrections do not contribute at very small
momentum. We will use this
result to show that the absolute value squared of the ground state
wave function of this state has the Halperin \cite{halperin} form at very long
distances, in the thermodynamic limit.

The $f$-sum rule can be derived as follows.
The retarded density and current correlation functions of this theory are, by
definition
\begin{equation}
D^{\alpha\beta }_{\mu \nu}(x,y)  = -i \theta (x_{0}-y_{0})
\times <G| [J^{\alpha}_{\mu}(x),J^{\beta}_{\nu}(y)] |G>
\label{eq:ret}
\end{equation}
where $J^{\alpha}_{\mu}$ ($\mu =0,1,2$) are the conserved  currents of the
theory, and $|G>$ is the ground state of the system.
Using this definition and the commutation relations between the currents,
one can derive the $f$-sum rule for the retarded density correlation
functions $D^{\alpha\beta }_{00}$. In units in which $e=c=\hbar =1$, it states
that
\begin{equation}
\int _{-\infty}^{\infty } \; {d\omega \over 2\pi} \; i \omega
       D^{\alpha\beta }_{00}(\omega,{\vec Q}) = {{\bar \rho}^{\alpha}
\over M}{\vec Q}^2  {\delta}^{\alpha\beta }
\label{eq:sum}
\end{equation}
This equation implies the conservation of the current in each
layer separately. It is easy to show that, in the basis of the total and
relative currents, {\it i.e.}, in the ${\bar \rho }_{\pm} =
{\bar \rho }^{1} {\pm}{\bar \rho }^{2}$ basis, eq~ (\ref{eq:sum}) states the
conservation of ${\bar \rho }_{+}$ and ${\bar \rho }_{-}$ independently.

On the other hand, it can be shown (see for instance reference \cite{book})
that the polarization tensor $K^{\alpha\beta }_{\mu\nu}$ and the density and
current correlation functions $D^{\alpha\beta }_{\mu\nu}$ satisfy the
following identity
\begin{equation}
K^{\alpha\beta }_{\mu\nu}(x,y) = - D^{\alpha\beta }_{\mu\nu}(x,y)
          + <{\delta J^{\alpha}_{\mu}(x) \over \delta A^{\beta}_{\nu}(y)}>
\label{eq:rel}
\end{equation}
Thus, eq~ (\ref{eq:sum}) is also valid if we replace
$D^{\alpha\beta }_{\mu\nu}$ by $K^{\alpha\beta }_{00}$ and we change the sign
in the r.h.s. of the equation.

\subsection{Ground state wave function for ($m,m,n$) states}
\label{subsec:waveunmedio}

We have found that for an ($m,m,n$) state, the leading order term in
${\vec Q}^2$ of the zero-zero component of the electromagnetic response is
given by
\begin{eqnarray}
K_{00}^{\alpha\beta } = &&- {{\bar \rho}\over 4M}\;
{{\vec Q}^2 \over {{\omega }^2 -{\omega }^2_c}
            + i \epsilon}  \left( \begin{array}{cc} 1& \;  1\\ 1& \; 1
                   \end{array}\; \right)  \nonumber \\
         && - {{\bar \rho}\over 4M}\;
{{\vec Q}^2 \over {{\omega }^2 -({(m-n) \bar \omega })^2}
            + i \epsilon}  \left( \begin{array}{cc} 1& \;  -1\\ -1& \; 1
                   \end{array}\; \right)
\label{eq:ksum}
\end{eqnarray}
Notice that the poles whose residues in $K_{00}^{\alpha\beta }$ are
proportional to ${\vec Q}^2$, have zero momentum frequency given by
$(m+n){\bar \omega}={\omega }_{c}$ and $(m-n){\bar \omega}$. In
particular, if $m=3$ and $n=1$, i.e., for the ($3,3,1$) state,
the expression in eq~(\ref{eq:ksum}) coincides with the result given by
eq~ (\ref{eq:resuno}) and ~(\ref{eq:restres}), provided that
${{\bar \rho}\over B} = {\nu \over 2\pi}$,
and that $ {( 1+ {16\over {\alpha}_{+}^{2}} )}^{-1}
                     + {( 1+ {16\over {\alpha}_{-}^{2}} )}^{-1} =1$,
with $\alpha _{\pm}$ defined by eq~(\ref{eq:unosub}).
In the gaussian approximation to the Chern-Simons Landau-Ginzburg theory
for the double layer systems \cite{wen,ezawa}, these two modes have also
residue proportional to ${\vec Q}^{2}$ in the density correlation function.
But in that approach, each of these modes separately saturates the $f$-sum
rule.

The correlation functions that we derive from the path integral formalism
are time-ordered. Therefore, if we use the relation between time-ordered and
retarded Green's functions, and eq~(\ref{eq:rel}) and ~(\ref{eq:ksum}),
we see that the leading
order term of $K_{00}^{\alpha\beta }$ saturates the $f$-sum rule,
eq ~(\ref{eq:sum}), already at the semiclassical level of our approach.
Notice that for this state ${\bar \rho}^{1}={\bar \rho}^{2}=
{{\bar \rho} \over 2}$.

Thus, the fermionic Chern-Simons approach gives the correct leading order
form for the density correlation function, in the sense that it is
consistent with the $f$-sum rule, at the semiclassical level of the
approximation.

It is important to remark that the coefficient of the leading order term of
$K_{00}^{\alpha\beta }$ can not be renormalized by higher order terms in the
gradient expansion, nor in the semiclassical expansion. In the case
of the gradient expansion, it is clear that higher order terms have higher
order powers of ${\vec Q}^2$, and then, do not modify the leading order term.
In the case of the corrections to $K_{00}^{\alpha\beta }$ originating in
higher order terms
in the semiclassical expansion, they also come with higher order powers of
${\vec Q}^2$. The reason of that is essentially the gauge invariance of the
system. This implies that the higher order correlation functions must be
transverse in real space, or equivalently they have higher order powers of
${\vec Q}^2$ in momentum space.
Being higher order terms in the ${\vec Q}^2$ expansion they can not change
the leading order term.

We will now follow the method used in reference \cite{l2} to
write the ground state wave function in the density representation.
We begin by recalling that the absolute value squared of the ground
state wave function in the density representation $|\Psi_0[\rho]|^2$ is
given by \cite{wavefnts}
\begin{equation}
 |\Psi_0[{\rho}_{1},{\rho}_{2}]|^2  = \int {\cal D} A_0^{\alpha} \;
 e^{-i\int d^{2}x \; A_0^{\alpha}({\vec x})\;\rho^{\alpha}({\vec x})}
 \lim_{A_{0}^{\alpha}(x)\to A_0^{\alpha}({\vec x})\delta (x_0)}
\langle 0|{T  e^{i\int d^3x\;A_{0}^{\alpha}(x)\;{\hat j}_0^{\alpha}(x)}}|0
\rangle
\label{eq:wf1}
\end{equation}
where ${\hat j}_0^{\alpha}(x)\equiv{\hat \rho}^{\alpha}(x)$.
The operators in this
expression are Heisenberg operators of the system in the absence of sources.
The vacuum expectation value in the integrand of eq~(\ref{eq:wf1}) can be
calculated from the generating functional of density correlation
functions, ${\cal Z}[{\tilde A}_{\mu}^{\alpha}]$ .

Clearly, we have
\begin{equation}
|\Psi_0[{\rho}_{1},{\rho}_{2}]|^2 = \int {\cal D} A_0^{\alpha}\;
    e^{-i\int d^{2}x \;A_0^{\alpha}({\vec x})\;
    \rho ^{\alpha}({\vec x})}
  \lim_{A_{0}^{\alpha}(x)\to A_{0}^{\alpha}({\vec x})\delta (x_0)} \;
     {\cal Z}[A_{0}^{\alpha},{\vec A}^{\alpha}=0] .
\label{eq:wf2}
\end{equation}
Eq~(\ref{eq:wf2}) tells us that $|\Psi_0[{\rho}_{1},{\rho}_{2}]|^2$
is determined by
the generating functional of equal-\-time density correlation functions.

The generating functional that appears in that expression is given by
\begin{equation}
 \lim_{A_{0}^{\alpha} (x)\to A_{0}^{\alpha} ({\vec x})\delta (x_0)} \;
{\cal Z}[A_{0}^{\alpha} ,{\vec A}^{\alpha} =0] =
  \int {{\cal D}{\psi^{*}}} {\cal D} \psi {\cal D} a_{\mu}^{\alpha} {\ }
  e^{i S(\psi^{*},\psi ,a_{\mu}^{\alpha} , A_{\mu}^{\alpha} ) }
\label{eq:wf3}
\end{equation}

The path integral on the r.h.s. of eq~(\ref{eq:wf3}) can be written in terms of
the effective action $S_{eff}(A_{\mu}^{\alpha})$ for the external
electromagnetic field.
We have seen that, in the thermodynamic limit, and for weak fields, the
effective action admits the expansion given by  eq~(\ref{eq:ems}).
Since we need only the density correlation functions, it suffices to know
the zero-zero component of $K_{\mu \nu}^{\alpha\beta}$.
In momentum space, and in the small ${\vec Q}^2$ limit, $K_{00}^{\alpha\beta}$
is given by eq~(\ref{eq:ksum}).
 We can see that the dominant
term in $K_{00}^{\alpha\beta}$ is of order $1/B$. Higher order terms in the
gradient expansion will contribute with higher powers of $1/B$. The same
observation applies for all the  corrections to $K_{00}^{\alpha\beta}$
originating in higher
order terms in the semiclassical expansion. Here the thermodynamic
limit is crucial since we are only taking into account fluctuations with
wavelengths short compared with the linear size of the system. The
higher order terms, which vanish like powers of ${\vec Q}^2/B$, can only
be neglected for an infinite system.

Using eqs~(\ref{eq:ems}) and (\ref{eq:ksum}), eq~ (\ref{eq:wf3}) becomes
\begin{equation}
 \lim_{A_{0}^{\alpha}(x)\to A_{0}^{\alpha}({\vec x})\delta (x_0)} \;
{\cal Z}[A_{0}^{\alpha},{\vec A}^{\alpha}=0]=
 e^{ {i\over 2} \int d^{2}x  d^{2}y A_{0}^{\alpha}({\vec x})
(\lim_{{x_0}\rightarrow {y_0}} K_{00}^{\alpha\beta}(x,y))
A_{0}^{\beta}({\vec y}) }
\label{eq:wf4}
\end{equation}
or, by Fourier transforming the exponent, we get
\begin{equation}
 \lim_{A_{0}^{\alpha}(x)\to A_{0}^{\alpha}({\vec x})\delta (x_0)} \;
{\cal Z}[A_{0}^{\alpha},{\vec A}^{\alpha}=0]=
e^{ {i\over 2} \int {d^{2}Q\over (2\pi)^2}{\ } A_{0}^{\alpha}({\vec Q})
        ({\int\limits_{-\infty}^{\infty } {d\omega \over{2\pi}}{\ }
  K_{00}^{\alpha\beta}(\omega, \vec Q)}){\ } A_{0}^{\beta}(-{\vec Q}) }
\label{eq:wf5}
\end{equation}

The terms dropped in the exponent of eq~(\ref{eq:wf4}) and
eq~(\ref{eq:wf5}) represent equal-\-time density  correlation functions with
more
than two densities. These terms give rise to three-\-body
corrections ( and higher) to the wave function and modify the Jastrow
form. The kernels of these non-linear contributions are,
by gauge invariance, required to be transverse. Thus, in
momentum space, the residues of their poles have higher powers in ${\vec
Q}^2$ than $K_{00}^{\alpha\beta}(\omega, \vec Q)$. Since, by dimensional
analysis,
each power of ${\vec Q}^2$ has to come with a factor of $1/B$, these
terms which are not bilinear in the densities are subleading
contributions in the limit $B \to \infty$. At the level of the
gaussian ( or semiclassical) approximation, these kernels are equal to
zero.
All of these considerations hold
provided that the Fourier transform of the pair potential
satisfies $ {\vec Q}^2 {\tilde V}(Q) \to 0$ as ${{\vec Q}^2 \to 0}$.

Replacing the expression for $K_{00}^{\alpha\beta}(\omega, \vec Q)$ given by
eq~ (\ref{eq:ksum}) into eq~ (\ref{eq:wf5}) and integrating out
$A_{0}^{\alpha}$, we obtain the following form for the absolute value
squared of the wave function
\begin{eqnarray}
|\Psi ({\vec x}_1,...,{\vec x}_{N_{1}},{\vec y}_1,...,{\vec x}_{N_{2}}) |^2& =&
           {\prod_{i<j=1}^{N_{1}}}{\ }|{\vec x}_i -{\vec x}_j|^{2m} \;
           {\prod_{i<j=1}^{N_{2}}}{\ }|{\vec y}_i -{\vec y}_j|^{2m} \;
     {\prod_{i=1}^{N_{1}}\prod_{j=1}^{N_{2}}}{\ }|{\vec x}_i -{\vec y}_j|^{2n}
\nonumber\\
  && {\ }{\exp }\big\{ -  {B\over 2} ({\sum_{i=1}^{N_{1}}}|{\vec x}_i|^2
                            + {\sum_{i=1}^{N_{2}}}|{\vec y}_i|^2 )\big\}
\label{eq:wf6}
\end{eqnarray}
where the coordinates ${\vec x}_i $ are in plane $1$, ${\vec y}_i $ are in
plane $2$, and $N_{1}=N_{2}= {N\over 2}$ for this state.

To get this result we have used that the eigenvalues of the local density
operator, in a Hilbert space with $N_{\alpha}$ particles, are
$\rho^{\alpha} (\vec x) =
{\sum_{i=1}^{N_{\alpha}}} {\delta (\vec x -{\vec x}_i)} - {\bar
\rho}^{\alpha}$.
A similar result was obtained recently by Schmeltzer and
Birman\cite{sch} who used a different approach.

Notice that the wave function of eq~ (\ref{eq:wf6}) is the absolute value
squared of the Halperin wave function \cite{halperin}. Numerical calculations
have established \cite{num} that this wave function accurately describes the
ground state wave function for the ($3,3,1$) state for $d=1.5 \ell_{c}$,
where $\ell_{c}$ is the cyclotron length.
We have shown that eq~(\ref{eq:wf6}) gives the exact form of the
ground state wave function at long distances and in the thermodynamic limit.
Since the {\it leading order term} of $K_{00}^{\alpha\beta}$
saturates the $f$-sum rule, higher order corrections in the expansion
cannot modify this result.

In the same way as for the single layer problem, we have  shown that,
in the thermodynamic limit, the exact asymptotic properties
of the wave function, when its arguments are separated by distances long
compared with the cyclotron length (but short compared with the linear
size of the system), are completely determined by the
long distance behavior of the equal-\-time density-\-density correlation
function ({\it i.e.}, the structure factor).

\subsection{Ground state wave function for ($m,m,m$) states}
\label{subsec:wavemm}

For these states, the leading order term in
${\vec Q}^2$ of the zero-zero component of the electromagnetic response is,
according to eq~(\ref{eq:resveinte}) and (\ref{eq:resveintiuno})
or (\ref{eq:restreintayuno})
\begin{eqnarray}
K_{00}^{\alpha\beta } = &&- {{\bar \rho}\over 4M}\;
{{\vec Q}^2 \over {{\omega }^2 -{\omega }^2_c}  + i \epsilon}
              \left( \begin{array}{cc} 1& \;  1\\ 1& \; 1
                   \end{array}\; \right)  \nonumber \\
         && - {{\bar \rho}\over 4M}\;
{{\vec Q}^2 \over {{\omega }^2 - v^{2}{\vec Q}^{2}
            + i \epsilon}  }
              \left( \begin{array}{cc} \;\; 1&  -1\\ -1& \;\; 1
                   \end{array}\; \right)
\label{eq:ksum2}
\end{eqnarray}
where we have used that ${{\bar \rho}\over B} = {\nu \over 2\pi}$, and that
$[ {( 1+ {16\over {\alpha}_{+}^{2}} )}^{-1}
                     + {( 1+ {16\over {\alpha}_{-}^{2}} )}^{-1}] =1$
with ${\alpha }_{\pm}$ defined by eq~(\ref{eq:veintiunosub}).

Following the same steps as in Sec~\ref{subsec:waveunmedio} we can prove that
the leading order term of $K_{00}^{\alpha\beta }$ saturates the $f$-sum rule,
eq ~(\ref{eq:sum}).  All the remarks about the exactness of this result
are also valid in this case. In this case too, the
Chern-Simons Landau-Ginzburg approach violates the $f$-sum rule at the
gaussian level \cite{wen,ezawa}.

Substituting eq~(\ref{eq:ksum2}) into the expression for the
 absolute value squared of the ground state
wave function (eq~(\ref{eq:wf2}) and (\ref{eq:wf5})) we obtain
\begin{eqnarray}
|\Psi ({\vec x}_1,...,{\vec x}_{N_{1}},{\vec y}_1,...,{\vec x}_{N_{2}}) |^2& =&
           {\prod_{i<j=1}^{N_{1}}}{\ }|{\vec x}_i -{\vec x}_j|^{2m} \;
           {\prod_{i<j=1}^{N_{2}}}{\ }|{\vec y}_i -{\vec y}_j|^{2m} \;
     {\prod_{i=1}^{N_{1}}\prod_{j=1}^{N_{2}}}{\ }|{\vec x}_i -{\vec y}_j|^{2m}
   \nonumber \\
&&   {\ }{\exp }\big\{ -  {B\over 2} ({\sum_{i=1}^{N_{1}}}|{\vec x}_i|^2
            + {\sum_{i=1}^{N_{2}}}|{\vec y}_i|^2 )\big\} \nonumber\\
&& {\exp } \big\{ -{m\; {v_{s}} \over {\omega _{c}}} \left(
{\sum_{i,j=1}^{N_{1}}} {1\over |{\vec x}_i -{\vec x}_j|}+
{\sum_{i,j=1}^{N_{2}}} {1\over |{\vec y}_i -{\vec y}_j|}-
2 {\sum_{i=1}^{N_{1}}} {\sum_{j=1}^{N_{2}}}{1\over |{\vec x}_i -{\vec y}_j|}
\right) \big\} \nonumber \\
\label{eq:wf8}
\end{eqnarray}
where the coordinates ${\vec x}_i $ are in plane $1$, ${\vec y}_i $ are in
plane $2$, and $N_{1}=N_{2}= {N\over 2}$ for this state.

Notice that this ground state wave function is not exactly the same as
the $(m,m,m)$  Halperin wave function, to which the true ground state
approaches as $d \rightarrow 0$. There is an extra
factor which comes from the fact that there is a gapless mode in the spectrum
of collective excitations.
This contribution is analogous to the phonon contribution to the wave function
of superfluid He$_4$, and just as in that problem, it is essential
to obtain the correct properties for the spatial correlations of the ground
state.
This contribution  is very small at long distances compared to the cyclotron
radius, and it can not be written only in terms of coordinates in the
lowest Landau level.

Based on the same arguments that we discussed in
Sec~\ref{subsec:waveunmedio}, we can argue here that eq~(\ref{eq:wf8}) is
an exact result for
the asymptotic form of the ground state wave function , at long distances
and in the thermodynamic limit.

\section{Hall conductance}
\label{sec:Hall}

We show now that, already within our approximation, this state does
exhibit the Fractional Hall Effect.
In order to do so, we will calculate the Hall conductance of the whole system.

Since we are only
interested in the leading long-distance behavior, it is sufficient to keep only
with those terms in the action of eq~(\ref{eq:gauss}) which have the smallest
number of derivatives, or
in momentum space, the smallest number of powers of $\vec Q$.
Therefore, the leading long distance
behavior ({\it i.e.}, small momentum)  of the effective action for the
fluctuations of the Chern-Simons gauge fields and of the
electromagnetic field is governed by the Chern-Simons term.
In this limit eq~(\ref{eq:gauss}) turns out to be
\begin{eqnarray}
S_{\rm eff} ({\tilde a}_{\mu}^{\alpha},{\tilde A}_{\mu}^{\alpha})
 && \approx i{{\bar \kappa}^{\alpha \beta} \over 2}
                       \int {d^{2}Q d\omega \over (2\pi)^3}
                       {\tilde a}_{\mu}^{\alpha} \;{\epsilon _{\mu\nu\lambda}}
                       { Q^{\lambda}}\; {\tilde a}_{\nu}^{\beta}
            -i{{\kappa}^{\alpha \beta} \over 2}
                       \int {d^{2}Q d\omega \over (2\pi)^3}
                       {\tilde a}_{\mu}^{\alpha}\; {\epsilon _{\mu\nu\lambda}}
                       { Q^{\lambda}}\; {\tilde A}_{\nu}^{\beta} \nonumber \\
        && -i{{\kappa}^{\alpha \beta} \over 2}
                       \int {d^{2}Q d\omega \over (2\pi)^3}
                       {\tilde A}_{\mu}^{\alpha}\; {\epsilon _{\mu\nu\lambda}}
                       { Q^{\lambda}}\; {\tilde a}_{\nu}^{\beta}
            +i{{\kappa}^{\alpha \beta} \over 2}
                       \int {d^{2}Q d\omega \over (2\pi)^3}
                       {\tilde A}_{\mu}^{\alpha}\; {\epsilon _{\mu\nu\lambda}}
                       { Q^{\lambda}}\; {\tilde A}_{\nu}^{\beta}
\label{eq:shal}
\end{eqnarray}
where $Q^{0}=\omega$ and $Q^{i}=-Q_{i}$ according with the convention that we
have used in reference \cite{l1}, and
${\bar \kappa}^{\alpha \beta}= {{p_{\alpha}}\over 2\pi}
{\delta}^{\alpha \beta} + {\kappa}^{\alpha \beta}$

The next step is to integrate the statistical gauge fields to obtain the
effective action for the electromagnetic field. In particular, we will need
to compute the inverse of the matrix ${\bar \kappa}^{\alpha \beta}$. This
inverse  only exists if
$\Delta = [({1\over p_{1}} + 2 s_{1})({1\over p_{2}} + 2 s_{2}) - n^{2}]
\not= 0$. Therefore we must consider two cases, the one in which
$\Delta \not= 0$ and the one in which $\Delta =0$.

\noindent Case $\Delta \not= 0$

Upon integrating over the statistical gauge fields in eq~(\ref{eq:shal}),
the effective action for the electromagnetic field results
\begin{equation}
S_{\rm eff}^{\rm em} ({\tilde A}_{\mu}^{\alpha})  \approx
i{{\kappa}^{\alpha \gamma} \over 2}[{\delta}^{\gamma \beta } -
                   ({\bar \kappa}^{-1})^{\gamma\delta} {\kappa}^{\delta \beta}]
                       \int {d^{2}Q d\omega \over (2\pi)^3}
                       {\tilde A}_{\mu}^{\alpha} \;
                       {\epsilon _{\mu\nu\lambda}}
                       { Q^{\lambda}}\; {\tilde A}_{\nu}^{\beta}
\label{eq:sel}
\end{equation}
where the coefficient ${\kappa}_{\rm eff}^{\alpha\beta}=
{\kappa}^{\alpha \gamma} [{\delta}^{\gamma \beta } -
({\bar \kappa}^{-1})^{\gamma\delta} {\kappa}^{\delta \beta}]$ is given by
\begin{eqnarray}
 {\kappa}_{\rm eff}^{\alpha \beta}
={1 \over 2\pi [({1\over p_{1}} + 2 s_{1})({1\over p_{2}} + 2 s_{2}) - n^{2}]}
\left( \begin{array}{cc}
({1\over p_{2}} + 2 s_{2})  & -n \\
-n & ({1\over p_{1}} + 2 s_{1})
\end{array}    \right)
\label{eq:sig}
\end{eqnarray}
In particular, if we consider the case in which both layers are coupled
to the same electromagnetic field, then ${\tilde A}_{\mu}^1 =
{\tilde A}_{\mu}^2={\tilde A}_{\mu}$, and the coefficient in the
effective action results
\begin{equation}
  {\kappa}_{\rm eff}\equiv \sum_{\alpha \beta}{\kappa}_{\rm eff}^{\alpha \beta}
={1 \over 2\pi} { {2n- ({1\over p_{1}} + 2 s_{1})-({1\over p_{2}} + 2 s_{2})}
\over  { n^{2}-({1\over p_{1}} + 2 s_{1})({1\over p_{2}} + 2 s_{2}) }}
={\nu \over 2\pi}
\label{eq:sigma}
\end{equation}
The electromagnetic current
$J_{\mu}$ induced in the system is obtained by differentiating the
effective action $S_{\rm eff} ({\tilde A}_{\mu})$ with respect
to the electromagnetic vector potential. The current is
$J_{\mu}={ {{\kappa}_{\rm eff}}\over 2} \epsilon_{\mu\nu \lambda}
{\tilde F}^{\nu\lambda}$.
Thus, if a weak external electric field ${\tilde E}_j$ is applied, the
induced current is
$J_k=  {\kappa}_{\rm eff} \epsilon _{lk} {\tilde E}_l $.
Therefore the coefficient $ {\kappa}_{\rm eff}$ is the {\it
actual} Hall conductance of the system.
\begin{equation}
 \sigma _{xy}\equiv  {\kappa}_{\rm eff} ={\nu\over 2 \pi}
\end{equation}
which is a {\it fractional} multiple of ${e^2\over h}$ (in units in
which $e=\hbar=1$). Thus, the uniform states exhibit a Fractional
Quantum Hall effect with the correct value of the Hall conductance.

\noindent Case $\Delta = 0$

It can be shown that when ${\bar \kappa}^{\alpha\beta}$ is not invertible,
i.e., it has a zero eigenvalue, the corresponding linear combination of
the gauge fields become massless. In other words, the {\it in phase} gauge
field ${\tilde a}_{\mu}^{+}=
{\tilde a}_{\mu}^{1}+{\tilde a}_{\mu}^{2}$ has a finite gap which couples
to the electromagnetic field  ${\tilde A}_{\mu}$, while the {\it out of phase}
gauge field ${\tilde a}_{\mu}^{-}=
{\tilde a}_{\mu}^{1}-{\tilde a}_{\mu}^{2}$ is gapless.

We will study in particular the case of the ($m,m,m$) states which satisfy the
condition $\Delta=0$.
For these states
\begin{eqnarray}
 {\bar \kappa}={-m  \over 2\pi (1-2m)}
\left( \begin{array}{cc}
1  & 1 \\
1 & 1
\end{array}    \right)
\label{eq:kapa}
\end{eqnarray}
which is clearly non invertible.

We can write the effective action for the fluctuations of the Chern-Simons
gauge fields (eq~(\ref{eq:shal})) in the basis defined by
${\tilde a}_{\mu}^{\pm} ={\tilde a}_{\mu}^{1}\pm {\tilde a}_{\mu}^{2}$ .
\begin{eqnarray}
S_{\rm eff} ({\tilde a}_{\mu}^{\alpha},{\tilde A}_{\mu}^{\alpha})
 && \approx {i\over 2} \int {d^{2}Q d\omega \over (2\pi)^3}
 [-{m\over {2\pi (1-2m)}} {\tilde a}_{\mu}^{+} \;{\epsilon _{\mu\nu\lambda}}
                       { Q^{\lambda}}\; {\tilde a}_{\nu}^{+}] \nonumber \\
        && -{i\over 2} \int {d^{2}Q d\omega \over (2\pi)^3}
 [{1\over {4\pi (1-2m)}}(-{\tilde a}_{\mu}^{+}{\epsilon _{\mu\nu\lambda}}
                       { Q^{\lambda}}\; {\tilde A}_{\nu}^{+}
                         + (2m-1)
                       {\tilde a}_{\mu}^{-}{\epsilon _{\mu\nu\lambda}}
                       { Q^{\lambda}}\; {\tilde A}_{\nu}^{-})] \nonumber \\
        && -{i\over 2} \int {d^{2}Q d\omega \over (2\pi)^3}
 [{1\over {4\pi (1-2m)}}(-{\tilde A}_{\mu}^{+}{\epsilon _{\mu\nu\lambda}}
                       { Q^{\lambda}}\;  {\tilde a}_{\nu}^{+}
                         + (2m-1)
                       {\tilde A}_{\mu}^{-}{\epsilon _{\mu\nu\lambda}}
                       { Q^{\lambda}}\; {\tilde a}_{\nu}^{-})] \nonumber \\
         && +{i\over 2} \int {d^{2}Q d\omega \over (2\pi)^3}
 [{1\over {4\pi (1-2m)}}(-{\tilde A}_{\mu}^{+}{\epsilon _{\mu\nu\lambda}}
                       { Q^{\lambda}}\; {\tilde A}_{\nu}^{+}
                         + (2m-1)
                       {\tilde A}_{\mu}^{-}{\epsilon _{\mu\nu\lambda}}
                       { Q^{\lambda}}\; {\tilde A}_{\nu}^{-})]
\label{eq:shal2}
\end{eqnarray}
The gauge field $ {\tilde a}_{\nu}^{-}$ appears as a Lagrange multiplier in
this action. The integration over it states that the current
${\epsilon _{\mu\nu\lambda}} { Q^{\lambda}}\; {\tilde A}_{\nu}^{-} $ vanishes.
This is trivially valid if the electromagnetic field is the same for both
layers, because $ {\tilde A}_{\nu}^{-}=0$.

The integration over $ {\tilde a}_{\nu}^{+}$ gives the effective action for the
field $ {\tilde A}_{\nu}^{+}$. If we consider the case in which
${\tilde A}_{\mu}^1 ={\tilde A}_{\mu}^2={\tilde A}_{\mu}$, the result is
\begin{equation}
S_{\rm eff}^{\rm em} ({\tilde A}_{\mu}) =
{i\over 2\pi m}    \int {d^{2}Q d\omega \over (2\pi)^3}
                       {\tilde A}_{\mu}       {\epsilon _{\mu\nu\lambda}}
                       { Q^{\lambda}}{\tilde A}_{\nu}
\label{eq:sel1}
\end{equation}
Following the same steps as in the previous case, the Hall conductance results
$\sigma _{xy}= {1\over 2\pi m}={\nu \over 2\pi}$ which is the correct value
for the ($m,m,m$) states.

\section{Quantum numbers of the quasiparticles}
\label{sec:stat}

In this section we evaluate the charge and statistics of the quasiparticles.
We will closely follow the methods and notation of reference \cite{book}.

We need to identify the operator which creates the quasiparticles within
the framework of the Chern-Simons theory. Let us consider the gauge invariant
operator which creates an excitation at the point ${\vec x}$ at $t=0$ in layer
$\alpha$, and destroys it at the point ${\vec y}$ at time $t=T$ in the same
layer, and which behaves as a quasihole
\begin{equation}
\psi^{*}_{\alpha}(x) e^{\big (-i \int_{\Gamma (x,y)} (a_{\mu}^{\alpha} +
A_{\mu}) dx_{\mu}\big )} \;  \psi_{\alpha}(y)
\label{eq:ch1}
\end{equation}
Here $\Gamma (x,y)$ is a path in space-time going from $(T,{\vec y})$ to
$(0,{\vec x})$. In this expression we are defining the quasihole operator in
the layer $\alpha$, therefore the index $\alpha$ is fixed, i.e., there is
no sum over $\alpha$ assumed. The operator in eq~(\ref{eq:ch1})
is invariant under gauge transformations of the statistical
gauge field, but depends on the choice of the path $\Gamma$. In this
expression, the fluctuations of the electromagnetic field ${\tilde A}_{\mu}$
have been switched off. The system only feels the uniform magnetic field
determined by $A_{\mu}$, and the statistical gauge fields.

Our goal is to evaluate the Green function $G_{\Gamma}(x,y)$ defined by
\begin{equation}
G_{\Gamma}(x,y) = \langle GS | T \; \big [
\psi^{*}_{\alpha}(x) e^{\big ( -i \int_{\Gamma (x,y)} (a_{\mu}^{\alpha} +
A_{\mu}) dx_{\mu}\big )} \;  \psi_{\alpha}(y)
\big ] |GS \rangle
\label{eq:gf}
\end{equation}
where $T$ is the time ordering operator.

We calculate this Green function in the path integral formalism, where it is
given by an average over the histories of the fermionic and statistical gauge
fields, weighted with the amplitude $\exp (i S(\psi^*,\psi,a_{\mu}^{\alpha},
A_{\mu}))$, where the action is defined by eq~(\ref{eq:pf}). After integrating
out the fermionic
fields, the Green function can be written, up to a normalization factor, as
\begin{equation}
G_{\Gamma}(x,y) =  \int {\cal D}a_{\mu}^{\alpha} \; G_{\alpha\alpha}(x,y)\;
e^ {-i \int_{\Gamma (x,y)} (a_{\mu}^{\alpha} +
A_{\mu}) dx_{\mu}} \; e^{i{\cal S}_{\rm eff}}
\label{eq:gf1}
\end{equation}
where ${\cal S}_{\rm eff}$ is given by eq~(\ref{eq:ese2}). The function
$G_{\alpha\alpha}(x,y)$ is the one particle Green function for a problem
of fermions in
a field determined by the statistical gauge field ($a_{\mu}^{\alpha}$) plus
the external magnetic field ($A_{\mu}$), and at finite particle density
determined by the chemical potential ($\mu_{\alpha}$). The one particle
Green function can be written in terms of a Feynman path integral as follows
\cite{feynman}
\begin{equation}
G_{\alpha\alpha} (x,y)= -i e^{i\mu T} \int {\cal D} {\vec z}[t]
\; e^{i{\cal S}[{\vec z}(t)]}
\label{eq:gf2}
\end{equation}
with the boundary conditions
\begin{eqnarray}
\lim _{t \rightarrow 0} z (t) &=& {\vec x} \nonumber \\
\lim _{t \rightarrow T} z (t) &=& {\vec y}
\label{eq:bc}
\end{eqnarray}
The weight $e^{(i\mu T)}$ serves to fix the number of particles. Since the
mean field solution has $p_{\alpha}$ effective Landau Levels filled,
the chemical potential has to be set to lie between the levels $p_{\alpha}$
and $p_{\alpha}+1$. The path integral in eq~(\ref{eq:gf2}) is a sum over all
the paths ${\tilde \Gamma}$ which go from $\vec x$ to $\vec y$ in time $T$. The
action is the standard one for non-relativistic particles coupled to a gauge
field
\begin{equation}
{\cal S}[{\vec z}(t)] = \int_{0}^{T} \big[ {M\over 2} ({d{\vec z}\over dt})^2
+ {dz^{\mu}\over dt}(t) (a_{\mu}^{\alpha} + A_{\mu}) \big]
\label{eq:gf3}
\end{equation}
Since there is an energy gap in this problem, in the long-distance, long-time
limit, the path integral is dominated by paths close to the solution of the
classical equations of motion. Therefore, the dominant trajectories are smooth.
Thus, it should be a good approximation to pull the integral over the
trajectories of the particles ${\vec z}(t)$ outside of the functional integral
over the statistical gauge fields. The integral over the trajectories
will be done at a later stage.
We can write the Green function as
\begin{equation}
G_{\Gamma} (x,y) \approx \int {\cal D} {\vec z}[t]
\; e^{i \int_{0}^{T}  {M\over 2} ({d{\vec z}\over dt})^2}
  \int {\cal D}a_{\mu}^{\beta} \;
e^ {-i \int_{\gamma } (a_{\mu}^{\alpha} + A_{\mu}) dx_{\mu}}
\; e^{i{\cal S}_{\rm eff}}
\label{eq:gf4}
\end{equation}
In this expression, the set of closed curves $\gamma$ represents paths which
are
the oriented sum of the path $\Gamma$ from $y$ to $x$ (which is fixed), and all
the possible paths $\tilde \gamma$ from $x$ to $y$ (which correspond to the
histories of the particles).

In the semiclassical approximation, the path integral over the statistical
gauge fields is replaced by an expansion around the solutions of the
classical equations of motion. In this approximation, the particle only
feels the electromagnetic field screened by the average of the statistical
gauge fields. In other words, the effective field felt by the particles is
$B_{\rm eff}^{\alpha}= B - ({\kappa}^{-1})^{\alpha\beta} {\bar \rho}_{\beta}$.
It is clear from eq~(\ref{eq:gf4}) that, for each closed trajectory $\gamma$
there is a constant factor which can be factored out from the functional
integral, and that corresponds to an Aharonov-Bohm phase factor for a
particle moving in the field  $B_{\rm eff}^{\alpha}$, not in the external
field $B$. In fact, the exponent of the Aharonov-Bohm phase factor is
equal to $B_{\rm eff}^{\alpha} A_{\perp}(\gamma)$, where $ A_{\perp}(\gamma)$
is the (spatial) cross sectional area bounded by the path $\gamma$. Defining
the effective charge as $q_{\rm eff}^{\alpha} =
{ B_{\rm eff}^{\alpha} \over B}$, and using eq~(\ref{eq:beff}), we find that
\begin{equation}
q_{\rm eff}^{\alpha} = { {\nu}^{\alpha} \over p^{\alpha} }
\label{eq:effq}
\end{equation}
In particular, for the ($m,m,n$) states, the effective charge in both layers
is the same and is given by $q_{\rm eff}= {1\over {(n+m)}}$.
This result coincides with the one in reference \cite{wz}.

The fractional statistics can be studied by considering the two particle
Green function. The generalization of the above formalism for this case is
straightforward. The only difference is that, for the two particle case, there
will be two sets of trajectories, one for each particle.
We will discuss first the case in which both particles are in the same layer,
let say, the layer $\alpha$. The two particle Green function
$G_{\Gamma}^{(2)}(x_{1},x_{2},y_{1},y_{2})$ is defined by
\begin{eqnarray}
&&G_{\Gamma}^{(2)}(x_{1},x_{2},y_{1},y_{2}) = \nonumber \\
&&\langle GS | T \big [
\psi^{*}_{\alpha}(x_{1}) \; \psi^{*}_{\alpha} (x_{2})
e^{\big ( -i \int_{\Gamma (x_{1},y_{1})} (a_{\mu}^{\alpha} +
A_{\mu}) \; dx_{\mu}
 -i \int_{\Gamma (x_{2},y_{2})} (a_{\mu}^{\alpha} +
A_{\mu}) \; dx_{\mu} \big )} \;  \psi_{\alpha}(y_{1})\psi_{\alpha}(y_{2})
\big ] |GS \rangle \nonumber \\
\label{eq:gf5}
\end{eqnarray}
Following the same steps as for the one particle case, we can write this two
particle Green function in terms of a path integral over the statistical gauge
fields and over the trajectories of the particles. In this case, the Grassman
integral automatically antisymmetrizes the two particle Green function, and it
comes as a sum of direct and exchange processes with the gauge fields
as a fixed background.
The two particle Green function turns out to be
\begin{eqnarray}
&&G_{\Gamma}^{(2)}(x_{1},x_{2},y_{1},y_{2}) \approx \nonumber \\
&& \int {\cal D} {\vec z}[t]
\; e^{i \int_{0}^{T}  {M\over 2} \sum_{j=1}^{2}({d{\vec z}_{j}\over dt})^2}
  \int {\cal D}a_{\mu}^{\beta} \; e^{i{\cal S}_{\rm eff}}
 \big[ e^ {-i \int_{\gamma_{\rm d} } (a_{\mu}^{\alpha} + A_{\mu}) dx_{\mu}}
 - e^ {-i \int_{\gamma_{\rm e} } (a_{\mu}^{\alpha} + A_{\mu}) dx_{\mu}} \big]
\label{eq:gf6}
\end{eqnarray}
where the path ${\gamma_{\rm d} }$ corresponds to direct processes (where
particle $1$ is destroyed at $x_{1}$ and created at $y_{1}$, and particle $2$
is destroyed at $x_{2}$ and created at $y_{2}$), and the
path ${\gamma_{\rm e} }$ corresponds to exchange processes (where particle
$1$ is destroyed at $x_{1}$ and created at $y_{2}$, and particle $2$
is destroyed at $x_{2}$ and created at $y_{1}$). Note that there is a relative
sign between these two processes.

In the low energy limit, the dominant paths are very long and wide. Therefore,
to compute the integral over the statistical gauge fields it will be
sufficient to consider  the effective action in the infrared limit. This
effective action only contains the Chern-Simons term, and it is given by
eq~(\ref{eq:shal}) but taking ${\tilde A}_{\mu}^{\alpha}=0$.
For both, direct and exchange processes, we need to calculate averages of the
form
$\langle \exp [i \int _{\gamma} dx^{\mu} a_{\mu}^{\alpha}]\rangle _{\rm CS}$,
where the subindex CS indicates that we only keep the Chern-Simons term in
the effective action. If the coefficient ${\bar \kappa}^{\alpha\beta}$
in the effective action is invertible, we can follow the steps described
in Appendix \ref{sec:AA} to calculate these averages. The result is
\begin{equation}
\langle e^{i \oint _{\gamma} dx^{\mu} a_{\mu}^{\alpha}}
\; \rangle _{\rm CS} =e^{{i \over 2}
{({\bar \kappa }^{-1})^{\alpha\alpha}}
\oint_{{\gamma}} d \sigma \; n_{\mu}^{\alpha} \; j_{\mu}^{\alpha}}
\label{eq:gf8}
\end{equation}
Recall that the index $\alpha$ is fixed and indicates the layer
to which the quasiparticle belongs.
The current $j_{\mu}^{\alpha}$ is a
three-vector of unit length tangent to the world lines and takes a non-zero
value {\it only} on the world lines of the particles ($\gamma $). The
integral $\oint_{\gamma} d \sigma \; n_{\mu}^{\alpha} \; j_{\mu}^{\alpha}$
 counts the number of times the current $j_{\mu}^{\alpha}$ pierces the
surface $\sigma$, therefore, it is equal to the linking number of the
curve $\gamma$, $\nu _{\gamma}$.
The configuration of paths can be classified according to their linking number.
The weights of the configurations with different linking numbers have different
phase factors. Also, configurations of paths from direct and exchange
processes also have different linking number. While the phase factors
themselves depend on the trajectories, and thus on the arbitrarily chosen paths
for the two particles, the relative phase only depends on the topological
properties of the configurations of paths and it is determined entirely by
the relative linking number $\Delta \nu_{\gamma}$. In particular we compare two
paths which form a linked knot with two paths which do not. In this case
$\Delta \nu_{\gamma}=1$.
Therefore, the difference of phase between the direct and the exchange terms
for these kind of paths in the two particle Green function is
\begin{equation}
\delta_{\alpha}= \pi \big( 1+ {(p_{\beta}+
                       { 2s_{\alpha}\over {4s_{1}s_{2}- n^{2}}   } )\over
{ (p_{1}+ { 2s_{2}\over {4s_{1}s_{2}- n^{2}}   } )
(p_{2}+ { 2s_{1}\over {4s_{1}s_{2}- n^{2}}   } )
- {n^{2}\over ({4s_{1}s_{2}- n^{2}})^{2}  }}} \big)
\label{eq:s1}
\end{equation}
where if $\alpha=1$, then $\beta=2$ and viceversa.
In particular, for the ($m,m,n$) states, the statistics is $\delta = -{m
\over {(m^2 -n^2)}}$, independent of the layer.
This result coincides with the one in reference \cite{wz}.

We define now the {\it relative} statistics of two particles
in different layers, as the relative phase factor that we obtain
if we compare the following processes
\begin{equation}
\langle GS | T \big [
\psi^{*}_{1}(x_{1}) \; \psi^{*}_{2}(x_{2})
 e^{\big ( -i \int_{\Gamma (x_{1},y_{1})} (a_{\mu}^{\alpha} +
A_{\mu}) \; dx_{\mu}
  -i \int_{\Gamma (x_{2},y_{2})} (a_{\mu}^{\alpha} +
A_{\mu}) \; dx_{\mu} \big )} \;  \psi_{1}(y_{1})\psi_{2}(y_{2})
\big ] |GS \rangle
\label{eq:gf10}
\end{equation}
where $x_1$ is a coordinate in plane $1$, and $x_2$ is a coordinate in plane
$2$, and
\begin{equation}
\langle GS | T \big [
\psi^{*}_{1}(x_{2}) \; \psi^{*}_{2} (x_{1})
e^{\big ( -i \int_{\Gamma (x_{2},y_{1})} (a_{\mu}^{\alpha} +
A_{\mu}) \; dx_{\mu}
 -i \int_{\Gamma (x_{1},y_{2})} (a_{\mu}^{\alpha} +
A_{\mu}) \; dx_{\mu} \big )} \;  \psi_{1}(y_{1})\psi_{2}(y_{2})
\big ] |GS \rangle
\label{eq:gf11}
\end{equation}
were $x_2$ and $x_1$ are the same coordinate as in eq~(\ref{eq:gf10}) but
living now in plane $1$ and $2$ respectively.

Following the same steps as above, we find that the {\it relative} statistics
$\delta _{12}$ is given by
\begin{equation}
\delta_{12}= \pi  {{n\over {4s_{1}s_{2}- n^{2}}} \over
{ (p_{1}+ { 2s_{2}\over {4s_{1}s_{2}- n^{2}}   } )
(p_{2}+ { 2s_{1}\over {4s_{1}s_{2}- n^{2}}   } )
- {n^{2}\over ({4s_{1}s_{2}- n^{2}})^{2}  }}}
\label{eq:s12}
\end{equation}
In particular, for the ($m,m,n$) states, the {\it relative} statistics
results $\delta ={n \over {(m^2 -n^2)}}$.

Up to this point, we have only considered the statistics for states such
that the matrix  $\kappa ^{\alpha\beta}$ is invertible. We consider now the
case in which this matrix is not invertible. In particular, we study the
($m,m,m$) states.
In order to calculate the two particle Green function for two particles
in different layers, we have to calculate
averages of the form
\begin{equation}
  \int {\cal D}a_{\mu}^{\gamma} \; e^{i{\cal S}_{\rm eff}}
  e^ {-i \oint_{\gamma_{\alpha} } a_{\mu}^{\alpha} j^{\mu}_{\alpha}
-i \oint_{\gamma_{\beta} } a_{\mu}^{\beta} j^{\mu}_{\beta}}
\label{eq:gf20}
\end{equation}
In this expression the indices $\alpha$ and $\beta$ are fixed, i.e., no sum
over them is assumed.
Since we are calculating the {\it relative} statistics, $\alpha =1$
and $\beta =2$ or viceversa. The currents $j_{\mu}$
have the same meaning as in eq~(\ref{eq:gf8}). In the low energy limit, we
can take again the effective action which only contains the Chern-Simons term,
and that is given by eq~(\ref{eq:shal}). In particular, for these states, it is
more convenient to work in the basis defined by $a_{\mu}^{\pm} = a_{\mu}^1
\pm a_{\mu}^2$. In this basis, the action is given by eq~(\ref{eq:shal2})
with ${\tilde A}_{\mu}^{\pm} =0$. Therefore, the expression in
eq~(\ref{eq:gf20}) results
\begin{equation}
  \int {\cal D}a_{\mu}^{+}{\cal D}a_{\mu}^{-} \;
\exp { {i\over 2} \int {d^{2}Q d\omega \over (2\pi)^3}
 [ {\bar \kappa}^{++}{\tilde a}_{\mu}^{+} \;{\epsilon _{\mu\nu\lambda}}
                       { Q^{\lambda}}\; {\tilde a}_{\nu}^{+}]
-i \int_{\gamma_{1}  \cup \gamma_{2}} a_{\mu}^{+} j^{\mu}_{+}
+  a_{\mu}^{-} j^{\mu}_{-}}
\label{eq:gf21}
\end{equation}
where ${\bar \kappa}^{++}=-{m\over {2\pi (1-2m)}}$.
The integral over $  a_{\mu}^{+}$ can be performed, and gives an expression
as the one in the r.h.s. of eq~(\ref{eq:gf8}) but replacing $\alpha$ by $+$.
The path integral $\oint d\sigma n^{+}_{\mu} j^{\mu}_{+}$ represents the
self-linking number of the trajectories defined by  $ j^{\mu}_{+}$ .

In the long wavelength limit, in which eq.~(\ref{eq:gf21}) is exact, the
gauge fields for the out of phase degrees of freedom, $a_{\mu}^-$, only
enters in the linear coupling to the external currents. Hence, in this
limit, the out of phase gauge field plays the role of a Lagrange
multiplier field. In particular,
the integral over $ a_{\mu}^{-}$ yields the constraint $ j_{\mu}^{-}=
  j_{\mu}^{1}-  j_{\mu}^{2}=0$. Since these currents are non-zero only on
the world lines of the particles, this constraint states that the particles
are forced to move together forming  a {\it bound state}.
This is a direct consequence of the existence of the gapless mode. If we
were to keep also terms which, in the effective action of the gauge
fields, vanish in the long wavelength limit, this
conclusion will be relaxed only a little. By dimensional analysis,
the effects of these terms should drop out like a power of a
length scale, which typically is the cyclotron length. The only
effect of these irrelevant terms is to give  the bound states  a
spacial extension of the order of that length scale. In other words, the
effect of the gapless mode
in the spectrum, is to induce a long-range interaction that forces the
quasiparticles in different layers to move together, forming a bound state.
Notice that this is a very stringent constraint since the ``constituent"
``independent" fermions from each layer have now been eliminated from
the spectrum by the existence of the gapless out phase gauge modes.
Indeed, the fact that there is a gapless mode which propagates with a
linear dispersion relation tells us that the first non-vanishing term in
the action of the out of phase gauge fields has a Maxwell-like form.
Thus, the Chern-Simons fermions feel an instantaneous force (mediated by
the gapless mode) which has a {\it logarithmic} dependence with the
distance. This is a confining force and leads to strong infrared
divergencies and to the {\it confinement} of the Chern-Simons fermions.
Only the bound states remain part of the physical spectrum. This picture
is strongly reminiscent of the mechanism by which the fermionic
states disappear from the spectrum of anyon
superfluids\cite{fhl,anyons}.

\section{Spin Singlet States and $SU(2)$ Symmetry}
\label{sec:su2}

In this section we discuss the application of Chern-Simons methods to a
2DEG which has an exact $SU(2)$ symmetry. In reference\cite{su2}
a non-abelian Chern-Simons approach was used to construct a theory of
the spin singlet $(m+1 , m+1, m)$ Halperin states. Here, we present
a generalization of the approach of reference\cite{su2} which will yield
the $SU(2)$ invariant hierarchies. A related, and completely equivalent
approach, was developed by Frohlich, Kerler and Marchetti~\cite{fkm}

In the approach of Balatsky and Fradkin\cite{su2} (BF), instead of the
two-component {\it abelian} Chern-Simons theory that we use in this
paper, a {\it non-abelian} Chern-Simons gauge field is introduced.
The advantage of the BF approach is that the $SU(2)$ invariance is
manifest and it is not the consequence of a subtle dynamical mechanism.
The disadvantage of the BF approach is that the non-abelian Chern-Simons
theory is substantially more sophisticated and technically more
demanding than the abelian theory that we use in the rest of this paper.
In the BF approach, the {\it
electron} is viewed as a composite object which is made of a particle
that carries the charge (the {\it holon}) and another particle that
carries the spin (the {\it spinon}). This {\it arbitrary} separation
gives rise to the existence of an abelian gauge symmetry (called {\it
RVB} by BF). The requirement of gauge invariance forces the holons and
spinons to be glued together in bound states, the electrons. In the BF
approach, the need of a non-abelian gauge field is a consequence
of the assignment of fractional (semion) statistics to both
holons and spinons. In this way,
$SU(2)$ fluxes are attached to a set of charge neutral,
spin-${\frac{1}{2}}$,
fermions, which become the spinons. The $SU(2)$ symmetry only admits
Bose, Fermi or semionic statistics. The holons, instead, are represented
by fermions attached to $U(1)$ fluxes. The $U(1)$ and $SU(2)$
Chern-Simons coupling constants must be chosen in such a way that the
holons and spinons are semions. Within this approach, the FQHE of the
spin-singlet states is the FQHE of the semions. The spin structure just
sits on top of the FQHE.

There is another, and more obvious, way to attach fluxes to particles
while keeping the $SU(2)$ spin symmetry untouched.
Belkhir and Jain\cite{jainspin} recently proposed a spin singlet wave
function for the
state with filling fraction $\nu={\frac{1}{2}}$, based on a composite
fermion picture. In their construction, they attach the {\it same}
number of pairs of flux quanta to {\it both} up and down spins (in other
words, they attach fluxes to the charge and not to the spin). The same
construction, but in the bosonic Chern-Simons language, was used before
by D.~H.~Lee and C.~Kane~\cite{leekane} and subsequently applied by
Sondhi, Karlhede, Kivelson and Rezayi~\cite{skkr} in their theory of
skyrmion states in polarized FQHE states. These approaches are
represented in
the $U(1) \otimes U(1)$ theory of the preceeding sections
by demanding that both up and down electrons see that same
flux at all times. This means to choose a
Chern-Simons matrix ${\kappa}^{\alpha \beta}$ which is proportional to the
identity, {\it i.~e.~}, $s_1=s_2$ and $n=0$. It is easy to use this
approach to get the $SU(2)$ limit of the Jain states, but with arbitrary
polarization. However, it is not a very efficient approach to get all of
the spin singlet states. The approach of BF deals with this states more
directly. It would be desirable to construct a theory which has as
particular cases both the $SU(2)$ theory of BF and the $U(1)$ theory of
reference ~\cite{leekane}. In what follows, we follow the BF approach.

More concretely, following BF, we introduce a holon field $\phi$ and a
spinon field $\chi_\alpha$ ($\alpha=\uparrow,\downarrow$) and represent
the {\it
electron} field operator $\psi_\alpha$ as $\psi_\alpha(x)=\phi(x) \;
\chi_\alpha(x)$. BF showed that, for an $SU(2)$ invariant system, the
system defined by the action of Eq.~(\ref{eq:ese1}) is equivalent to the
following theory of (interacting) spinons and holons. Let the total
action $\cal S$ be the sum of a charge, spin and interaction terms
\begin{equation}
{\cal S}= {\cal S}_{\rm charge}+ {\cal S}_{\rm spin} + {\cal S}_{\rm
interaction}
\end{equation}
where the action for the charge degrees of freedom is
\begin{equation}
{\cal S}_{\rm charge}=\int d^3x\;\left( \phi^{\dagger}(x)\; (
iD_0^c+\mu)
\; \phi(x)+{\frac{1}{2M}} \phi^{\dagger}(x)\; {\vec D_c}^2 \phi(x)
\right)+\int d^3x\;{\frac{\theta}{2}} \epsilon_{\mu \nu \lambda}
a^\mu(x) \partial^\nu a^\lambda(x)
\label{eq:scharge}
\end{equation}
while the action for spin is
\begin{eqnarray}
{\cal S}_{\rm spin}=&\int& d^3x\;\left( \chi_\alpha^{\dagger}(x)\;
iD_0^s
\; \chi_\alpha(x)+{\frac{1}{2M}} \chi_\alpha^{\dagger}(x)\; {\vec D_s}^2
\chi_\alpha(x)
\right)\nonumber\\
-&\int& d^3x \; {\frac{k}{4 \pi}} \epsilon_{\mu \nu
\lambda}
{\rm tr}( b^\mu(x) \partial^\nu b^\lambda(x)+{\frac {2}{3}} b^\mu
(x) b^\nu (x) b^\lambda (x))
\label{eq:sspin}
\end{eqnarray}
and an (instantaneous) pair interaction term
\begin{equation}
{\cal S}_{\rm interaction}=-\int d^3x \; \int d^3x' \; {\frac{1}{2}}
(\rho(x)-{\bar \rho}) V(x-x') (\rho(x')-{\bar \rho})
\label{eq:sint}
\end{equation}
In equation (\ref{eq:scharge}), $a_\mu$ is the statistical vector
potential
which turns the holons into semions. This condition requires that the
$U(1)$ Chern-Simons coupling constant $\theta$ be restricted to the
values
\begin{equation}
{\frac{1}{\theta}}=-{\frac{2\pi}{m}}+2\pi 2s
\label{eq:thetasemion}
\end{equation}
where semion statistics requires that $m=\pm 2$ and $s$ is an arbitrary
integer. Likewise, in eq.~(\ref{eq:sspin}) $b_\mu$ is the $SU(2)$
non-abelian statistical gauge field which takes values on the $SU(2)$
algebra. Hence, we can expand the field  in the form
$b_\mu(x)=b_\mu^a(x)\; \tau^a$, where
$\tau^a$ ($a=1,2,3$) are the three generators of $SU(2)$ in the spinor
representation , {\it i.~e.~} the set of $2 \times 2$ Pauli matrices.
The $SU(2)$ Chern-Simons coupling constant $k$, the {\it level}
of the Chern-Simons theory, for our system, is equal to\cite{su2}
$k=\pm
1$. Hence, we have a (``trivial") level one Chern-Simons theory. All the
representations of a level one Chern-Simons theory are known to be
abelian and to correspond to abelian fractional statistics of fermions,
bosons or semions\cite{wittencs}. The only ambiguity left is the sign of
$k$ which is the chirality (or handedness) of the semion. There is a
similar sign ambiguity in the coupling constant $\theta$ in
eq.~(\ref{eq:thetasemion}). Different choices of these signs lead to
different statistics for the bound states of holons and spinons. The
requirement that the bound state be an electron, which is a fermion,
leads to the condition ${\rm sign}(k)={\rm sign}(m)$.

The $U(1)$ and $SU(2)$ charge and spin
covariant derivatives are
\begin{eqnarray}
D_\mu^c&=&\partial_\mu-i(A_\mu+a_\mu+c_\mu)\nonumber\\
D_\mu^s&=&\;I\;\partial_\mu-i(b_\mu-I\;c_\mu)
\label{eq:covder}
\end{eqnarray}
where $A_\mu$ is the external electromagnetic field and $I$ is the $2
\times 2$ identity matrix. The gauge field $c_\mu$ is the
$U(1)$ ``{\em RVB}" gauge field which glues spins and charges together.
The covariant derivatives have been chosen in such a way that
holons and spinons have opposite charge with respect to $c_\mu$. Hence,
the strong fluctuations of this field binds holons and spinons into
states which are locally singlets under the ``RVB" gauge
transformations, {\it i.~e.~}, electrons. In fact, all the gauge field
$c_\mu$ does is to enforce the constraint that the 3-current
of the holons equals the 3-current of the spinons, as an operator
statement in the physical Hilbert space. This is seen clearly from the
equation of motion generated by $c_\mu$
\begin{equation}
J^{\rm RVB}(x)=\frac{\delta {\cal S}}{\delta c_0(x)}=0 \;
\Longrightarrow\;\phi^{\dagger}(x)\phi(x)-\sum_{\sigma=\uparrow,
\downarrow}\chi_\sigma^{\dagger}(x)\chi_\sigma(x)=0
\label{eq:jrvb}
\end{equation}
Thus $N_{\rm e}$, the number of charges, and $N_{\uparrow}$ and
$N_{\downarrow}$, the number of up and down spins, must obey the obvious
relation $N_{\rm e}=N_{\uparrow}+N_{\downarrow}$. Also,
eq.~(\ref{eq:jrvb}) tells us that the local particle density
operator $\rho(x)$ can be identified with the holon charge density
operator, {\it i.~e.~}, $\rho(x)= \phi^{\dagger}(x)\phi(x)$.

We will not attempt to go into the details of the full non-abelian
theory. Rather, we will use it to determine the allowed fractions for
$SU(2)$ invariant states. Thus, we will just consider the AFA equations
for this theory. There are two sets of AFA equations, one for the charge
sector and one for the spin sector. The AFA equations for the charge
sector are just the AFA equations for a charged interacting semion
liquid with $N_{\rm e}$ particles in an uniform magnetic field with
$N_\phi$ flux quanta, {\it i.~e.~}, a FQHE of semions. A simple
application of the methods of reference\cite{l1} yields the constraint
for the charge density operator
\begin{equation}
\frac{\delta {\cal S}}{\delta a_0(x)}=0 \;\Longrightarrow \;
j_0(x) =-\theta {\cal B}_c (x)
\label{eq:semioncharge}
\end{equation}
where $j_0(x) $ is the charge density operator. This equation, when
specialized on fluid states, for which
 ${\bar \rho}$ and ${\cal B}_c(x)$ are
the average density and average $U(1)$ statistical charge flux
respectively, yields the allowed filling fractions.
Similar considerations for the $SU(2)$ gauge
field $b_\mu^a$ (with $a=1,2,3$) yield a constraint for the local spin
density operator
\begin{equation}
{\frac{\delta {\cal S}}{\delta b_\mu(x)^a}}=0 \; \Longrightarrow \;
j_0^{a}(x)=
\psi_\sigma^{\dagger} \tau^a \psi_\sigma=
{\frac{k}{2 \pi}}{\cal B}_s^{a}
\label{eq:semionspin}
\end{equation}
where ${\cal B}_s^{a}$ is the $SU(2)$ spin flux.

For fluid states, we generalize the Average Field
Approximation and replace these exact local operator
identities by translationally invariant averages. This replacement may
be problematic in the case of the $RVB$ gauge field since its only
mission is to enforce {\it exactly} the constraint eq.~(\ref{eq:jrvb}).
At the level of the wave functions for the allowed ground states, this
constraint simply means that every coordinate for a charge degree of
freedom has
to coincide with the coordinate of a spin degree of freedom.

Thus, we seek fluid states with $N_e$
particles, $N_\phi$ flux quanta and filling
fraction $\nu=N_e/N_\phi$. We wish to determine the filling
fractions and spin for which the ground state is a fluid.
In addition
to the usual AFA equation
for the charge, that will give the allowed fractions, we will
now get conditions for the spin and polarization of the allowed states.

In the charge sector, we have a FQHE of spinless semions. In this
case, the AFA
consists of  a system of $N_e$ spinless fermions filling up
effective Landau levels, exactly as in our earlier work for spin
polarized electrons (and anyons!)\cite{l1}. The same line of argument
that was used
to derive the allowed fractions in Section \ref{subsec:meanf} now tells
us that, from eq.~(\ref{eq:semioncharge}), the effective number of
fluxes is
\begin{equation}
{\bar N}_{\phi}=N_{\phi}-{\frac{N_e}{2 \pi \theta}}
\label{eq:fluxsu2}
\end{equation}
and, hence,
the allowed fractions, $\nu^{\pm}$, for the $SU(2)$ fluid states satisfy
\begin{equation}
{\frac{1}{\nu^{\pm}}}=\pm {\frac{1}{p}}-{\frac{1}{2 \pi \theta}}
\label{eq:su2frac}
\end{equation}
where $p$ is a positive integer and the $+$ sign corresponds to a
``particle-like" FQHE  ({\it i.~e.~}, the effective flux parallel to the
external flux)
while the $-$ sign holds for a ``hole-like" FQHE ({\it i.~e.~}, the
effective flux anti-parallel to the external flux) . By using the allowed
values of $\theta$ we find that the allowed fractions for the
$SU(2)$ fluid states are of the form
\begin{equation}
\nu^{\pm}(p,s;m)={\frac{mp}{(2sm-1)p \pm m}}\equiv{\frac{2p}{\pm 2 +
(4s-{\rm sign}(m))p}} \label{eq:allowed}
\end{equation}
where we have specialized for the case of interest, $m=2 {\rm sign}(m)$.

The hierarchy of FQHE states of eq.~(\ref{eq:allowed}) is a
generalization of the states found
by BF, which are obtained by setting $p=+1$ and $m=+2$, {\it i.~e.~},
$\nu^+(1,s;-2)={\frac{2}{4s+1}}=2,{\frac{2}{5}},{\frac{2}{9}}, \ldots$ ,
which coincide with the Halperin-Haldane spin singlet states.

The state
with $\nu={\frac{1}{2}}$
is also part of the hierarchy of eq.~(\ref{eq:allowed}), where it is
realized as the state with $p=2$ ($-2$) for $m=+2$($-2$) and $s=1$.
Numerical
studies show that, for Coulomb-like interactions, this state is not
favored and that the $(3,3,1)$ is an accurate representation of the
ground state.

Finally, the this hierarchy has a state at filling fraction
${\frac{5}{2}}$. This state is found as the $p=-10$ (``hole-like"),
$m=-2$, $s=0$, member of the $SU(2)$ hierarchy, or as the $p_1=p_2=-5$,
$s_1=s_2=s$ and $n+2s=1$ member of the $U(1) \otimes U(1)$ hierarchy.
This
is the only spin singlet state that has yet been seen experimentally.

In contrast with the states that we found in section \ref{subsec:meanf}
using the abelian theory for bilayers,
in the $SU(2)$ theory, the states {\it at each filling fraction}, are
arranged in irreducible representations (multiplets) of $SU(2)$.
Thus, the $SU(2)$ theory does not have any redundant states.
The spin and polarization of the states is determined by
eq.~(\ref{eq:semionspin}). The only subtlety here is that, since the
components of the total spin do not commute with each other, one can
only determine the total spin and total projection. For a system without
a boundary, the choice of total spin $S$ and total projection along an
arbitrary polarization axis, say $S_z$, are constants of motion which
are invariant under local $SU(2)$ gauge transformations (but, of course,
change under global $SU(2)$ rotations). There are two generic situations
of physical interest: (a) spin {\it singlet} states (or with {\it
microscopic} total spin $S/N_e \approx O(1/N_e)$) and (b) states with
macroscopic spin, $S \approx N_e$, {\it i.~e.~} {\it ferromagnetic}
states.
Thus, the total z-component of the spin polarization
$M={\frac{1}{2}}(N_{\uparrow}-N_{\downarrow})$ obeys
\begin{equation}
2M=N_{\uparrow}-N_{\downarrow}={\frac{k}{2 \pi}} \langle
{\cal B}_3^s \rangle L^2
\label{eq:allowedspin}
\end{equation}
where $L^2$ is the area. It is clear that it is possible to construct
all multiplets with spin $|S| \leq {\frac{N_e}{2}}$. In the
thermodynamic limit, the spin {\it singlet} states have $S=0$ and,
hence, $N_{\uparrow}=N_{\downarrow}=N_e/2$. In contrast, the
ferromagnetic states have, with an appropriate choice of the
quantization axis, a non-vanishing {\it extensive} value of $M$ and,
hence, a non-zero value of ${\cal B}_3^s$.

The wave functions for the spin sector of the spin singlet states have
to be determined from the states of a level one $SU(2)$ Chern-Simons
gauge theory with $N_e$ sources in the fundamental representation. It
was shown by Witten\cite{wittencs} that these wave functions are
correlation functions of conformal blocks of a conformal field theory in
two Euclidean dimensions, the $SU(2)$ level one Wess-Zumino-Witten
model. This fact was used by Read and Moore\cite{readmoore} and by
Balatsky and Fradkin\cite{su2} to show that, the wave function of the
spin singlet FQHE has a factor which is precisely this conformal block
correlation function. It was also noticed\cite{su2} that this factor
coincides with the Kalmeyer-Laughlin wave function\cite{kl} for a
Spin Liquid.

The states with macroscopic spin have a somewhat different physics. The
existence of a non-zero average field should make a mean field approach
more sound. The spin sector of this mean field theory has
$N_{\uparrow,\downarrow}={\frac {1}{2}}N_e \pm M$ spin up and spin down
spinons each feeling an effective uniform magnetic field of $\pm
{\frac{2 \pi}{k}}M$. Notice that, because of the $SU(2)$ invariance,
there is no Zeeman term and only the orbital degrees of freedom see
this spin-dependent external field. It is easy to see that the highest
weight ferromagnetic state with maximal spin is obtained by filling up
the lowest Landau level of the up spins while leaving the down spin
sector empty.

The charge and spin sectors are not decoupled from each other.
Firstly, the constraint of eq.~(\ref{eq:jrvb}) sets the {\it local}
charge density to be the same as the {\it local} spin density. The wave
functions of the allowed states have to satisfy this {\it local}
property. Secondly, if the system is $SU(2)$ invariant, all of the
states {\it in a given} $SU(2)$ multiplet must have the {\it same}
filling fraction. Since the {\it fully polarized} states have to span
all of the Jain states for a single layer system, we must conclude that
$SU(2)$ states which are not in a main Jain hierarchy cannot achieve the
maximum polarization. In other terms, there is an upper bound for the
spin polarization and, hence, for the total spin itself. Thus, the
filling fraction $\nu$ and the spin $S$ of the state cannot be set
completely independent from each other for the allowed states. In other
terms, there should exist  a set of selection rules which determine the
allowed combinations of total spin and filling fraction.
Similarly, it should be possible to construct a unified theory of
all the FHQE states with $SU(2)$ symmetries, instead of the apparently
separate descriptions of the spin singlet and the fully polarizable
states that
we use here. We will return to these issues in a separate publication.

We conclude this section with a comparison of the states that are
obtained by this $SU(2)$-symmetric approach and the $U(1) \otimes U(1)$
theory that we use in the rest of this paper. A direct inspection of the
allowed fractions
eq.~(\ref{eq:fillfrac}) and eq.~(\ref{eq:allowed}) for the $U(1) \otimes
U(1)$
and $SU(2)$ theories respectively, shows that they do not yield the same
allowed fractions. For instance, the ``Fermi Liquid" (compressible)
states, with the same occupancy of the two layers, allowed by
eq.~(\ref{eq:fillfrac}) have filling fractions
${\frac{2}{r}}=2,1,{\frac{2}{3}},
{\frac{1}{2}}, {\frac{2}{5}}, {\frac{1}{3}}, {\frac{2}{7}},
{\frac{1}{4}}, {\frac{2}{9}}, \ldots$ ($r=1,2, \ldots$). In contrast,
the allowed $SU(2)$ ``Fermi Liquid" (compressible) states are
${\frac{2}{4s \pm 1}}=2,{\frac{2}{5}}, {\frac{2}{3}}, {\frac{2}{9}},
{\frac{2}{7}},\ldots$. Clearly, the fractions ${\frac{1}{k}}$ (with
$k=1, 2,3,\ldots$) cannot be realized as $SU(2)$ compressible states.
Among the states which appear in both hierarchies, we find an
incompressible spin unpolarized state at filling fraction
${\frac{4}{11}}$. It is worth noting, that there is experimental
evidence of an incompressible state at ${\frac{4}{11}}$. It is not
possible to construct a fully polarized Jain state with this filling
fraction although it may be constructed as a hierarchical state. It is
strange that this is the only observed fraction for which a hierarchical
construction is needed. It is believed that the experimentally observed
state is polarized.

We have further checked that
all of the states in levels $1$ and $2$ in the
$SU(2)$ hierarchy span the entire level $1$ $U(1) \otimes U(1)$ states.
Similarly, the level $4$ $SU(2)$ states span the level $2$  $U(1)
\otimes U(1)$ states, and the levels $3$ and $6$ $SU(2)$ states span the
level $3$ $U(1) \otimes U(1)$ states. However,  a large
number of incompressible $U(1) \otimes U(1)$
states cannot be realized as $SU(2)$ states. In a way, this should not
be surprising since the $U(1) \otimes U(1)$ symmetric theory may only
generate $SU(2)$ as a dynamical symmetry. Nevertheless, it is a puzzling
result since, in this discussion, the form of the interaction terms
has not entered and, hence, the symmetries of the Hamiltonian have not
had a chance to play
any role yet. However, even more surprising is the fact that not all of
the
$SU(2)$ states can be realized as $U(1) \otimes U(1)$ states. It is easy
to check, for instance, that the allowed $SU(2)$ state with
$\nu={\frac{10}{7}}$ has no counterpart in the $U(1)
\otimes U(1)$ states (unless polarized states are also considered).

\section{Conclusions}
\label{sec:conc}

In this work, we generalized the fermionic Chern-Simons
theory for the Fractional Hall Effect (FQHE) which we developed before, to the
case of double layer-systems.

We studied the semiclassical approximation around the liquid-like mean field
solutions.
We found that we can describe a hierarchy of double-layer FQHE states which
include the ($m,m,n$) states, and the ($m,m,m$) states, with filling fractions
$\nu={1\over {m+n}}$ and $\nu = {1\over m}$ respectively. For all these
states, the mean field ground state has a gap to all the excitations and the
semiclassical approximation is a well controlled perturbative expansion.
Within the solutions of the saddle point equations, we also encounter the
generalization to the double-layer systems, of the compressible states
discussed by Halperin et al \cite{hlr}. It turns out that the liquid-like
solutions of the saddle point equations predict a very rich phase
diagram for double-layer systems. The structure of this phase diagram
should depend on the particular microscopic form of the pair
interactions.

We studied the electromagnetic response functions of these systems and
calculated the experimentally accessible optical properties.
In general, for the so called $(m, m, n)$ states, we found that the
spectrum of collective excitations has a gap for all the modes.
We derived the absolute value squared of the ground state wave function.
We found that it has the Jastrow-Slater form, with the exponents determined
by the coefficients $m$, and $n$.
We also found that the  $(m,m,m)$ states,
have a gapless mode which may be
related with the appearance of interlayer coherence\cite{wen,ezawa}.
Our results also indicate that the gapless mode makes a contribution to
the wave function of the $(m,m,m)$ states analogous to the phonon
contribution to the wave function of superfluid $\rm{He}_4$. This factor
is crucial to obtain the correct spatial correlations from the ground state
wave function.

In all the cases we verified that the density correlation function
saturates the $f$-sum rules associated with the conservation of the
number of particles in each layer separately.

We calculated the Hall conductance and verified that it gives the correct
value already at the semiclassical level of our approach.

We calculated the charge and statistics of the quasiparticles. We saw that the
charge is determined by the filling fraction in the layer divided by the
corresponding number of effective Landau levels filled.
We found that the statistics of the quasiparticles
is well defined only for states
which satisfy that the Chern-Simons matrix of the effective action for the
statistical gauge fields (${\bar \kappa }^{\alpha\beta}$) is invertible. For
instance, for the ($m,m,n$) states, the statistics for
a quasiparticle in a given plane is determined by $m\over {m^2 - n^2}$,
and  the {\it relative} statistics is given by  $n\over {m^2 - n^2}$.
On the other hand, for states such that ${\bar \kappa }^{\alpha\beta}$ is
not invertible, as for instance the ($m,m,m$) states, we found that
the statistics of the quasiparticles is not well defined.
In particular, we saw that the effect of the
gapless mode is to induce a long-range attraction that forces the
quasiparticles in different layers to move together, forming a bound state.

We have also compared the results of this $U(1) \otimes U(1)$ theory
with a generalization of the $SU(2)$ theory of Balatsky and Fradkin. We
showed that, for many fractions, the $U(1) \otimes U(1)$ can be embedded
inside an $SU(2)$ theory. In this way we were able to identify $SU(2)$
states in our theory for bilayers. However, we also found that the two
hierarchies are not completely equivalent.

Finally, we would like to remark that, within our approach, the gap of the
collective modes is always proportional to
the effective cyclotron frequency. In particular, this holds also for
the {\it out
of phase} modes. In principle, one expects that the zero momentum frequency of
these modes should be determined essentially by the interlayer pair potential,
$V_{12}$, but, as in the single layer problem, we can not obtain this result
within the gaussian approximation and we expect the non-gaussian
corrections to give a substantial correction to this energy gap as well.

\section{Acknowledgements}

One of us (A.L.), thanks Shivaji Sondhi for helpfull discussions. EF is
also grateful to Jainandra Jain for an enlightening discussion of his
work. EF wishes to thank the Aspen Center of Physics for its kind
hospitality. This work was supported in part
by the National Science Foundation through the grant NSF DMR-91-22385 at
the University of Illinois at Urbana-Champaign (EF),
by the American Association of University Women
(AL), by the Science and Technology Center for Superconductivity at the
University of Illinois at Urbana-Champaign and by the Research Board of
the University of Illinois at Urbana-Champaign.

\newpage

\appendix
\section{}
\label{sec:AA}

In this appendix we follow the method described in Appendix A of reference
\cite{l1} to prove that both theories yield the same physical amplitudes
(i.e., that they are equivalent).
The idea is to compute the same (arbitrary) amplitude in both schemes.
We do the calculation in the path-integral language. We can follow step by step
what we did in reference \cite{l1}. The only difference is that in this
case, there
are two species of fermions. The fermions living in one layer are
identical, but they are distinguishable from the fermions living in the other
layer.

The standard action for nonrelativistic particles coupled to the
electromagnetic field is
\begin{equation}
{\cal S} =   \sum _{\alpha}\int_{-\infty}^{+\infty}dt \int d^{2}z \;
            j_{\mu}^{\alpha}({\vec z}, t) \; A^{\mu}_{\alpha}
          + {\cal S}_{\rm matter}
\label{eq:ac1}
\end{equation}
where $\mu =0,1,2$ is the space-time index, $\alpha =1,2$ labels the layer,
and ${\cal S}_{\rm matter}$ includes both the kinetic energy of the
particles as well as their pair interactions. The boundary
conditions are \begin{eqnarray}
\lim _{t \rightarrow -\infty} z_{j} (t) &=& x_{j} \nonumber \\
\lim _{t \rightarrow +\infty} z_{j} (t) &=& x_{Pj}
\label{eq:zeta}
\end{eqnarray}
where the final states are a permutation of the initial states.
The current $j_{\mu}^{\alpha}({\vec z},t)$ is a
three-vector of unit length tangent to the world lines and takes a non-zero
value {\it only} on the world lines of the particles ($\Gamma (\alpha )$)
\begin{eqnarray}
j_0^{\alpha}({\vec z},t)&=&\sum_{j=1}^{N_{\alpha}} \delta
   ({\vec z}_j(t)-{\vec z}(t)) \nonumber \\
      {\vec j}^{\alpha }({\vec z},t)&=&\sum_{j=1}^{N_{\alpha}} \delta
       ({\vec z}_j(t)-{\vec z}(t)) {d {\vec z}_j\over dt}
\label{eq:jota}
\end{eqnarray}

If we couple the above system to the Chern-Simons fields $a_{\mu}^{\alpha}$,
the action becomes
\begin{equation}
{\cal S}_{\rm new} = {\cal S} + \sum _{\alpha}\;
        \int_{-\infty}^{+\infty}dt \int d^{2}z \;
              j_{\mu}^{\alpha}({\vec z}, t)\; a^{\mu}_{\alpha}
          +  \sum _{\alpha\beta}\; {\kappa _{\alpha \beta}\over 2}
\int d^3z\;
\epsilon_{\mu \nu \lambda} {a}^{\mu}_{\alpha} \partial {a}^{\mu}_{\beta}
\label{eq:ac2}
\end{equation}
It is possible to perform the functional integral over the statistical gauge
fields exactly. These fields enter in only two terms of $S_{\rm new}$: the
current term and the Chern-Simons term. Thus, the average over all
the configurations of the statistical gauge fields has the form
\begin{equation}
\langle \exp(i \sum _{\alpha}\; \int d^3 z j_{\mu}^{\alpha}(z)
a^{\mu}_{\alpha }(z)) \rangle_{\rm CS}\equiv \exp(i I[ j_{\mu}^{\beta } ])
\label{eq:pi}
\end{equation}
where the notation $\langle O \rangle _{\rm CS}$ indicates the average of
the operator $O$ over the statistical gauge fields with only a Chern-Simons
action.

Our goal is to find out which has to be the form of the Chern-Simons
coefficient, i.e., of the matrix $\kappa _{\alpha\beta}$, for the r.h.s. of
eq~(\ref{eq:pi}) to be equal to one.

After integrating out the Chern-Simons fields we obtain the following
expression for $I[ j_{\mu}^{\beta } ]$
\begin{equation}
I[ j_{\mu}^{\beta } ] = \sum _{\alpha\beta}\;
{ {({\kappa }^{-1})^{\alpha\beta}} \over 2}
 \int d^3z \int d^3z' j_{\mu}^{\alpha}(z) G^{\mu \nu}(z,z')j_{\nu}^{\beta}(z')
\label{eq:ii}
\end{equation}
where $G_{\mu \nu}(z,z')$ is the Green function
\begin{equation}
G_{\mu \nu}(z,z')= \epsilon_{\mu \nu \lambda}
\partial_{\lambda}^{(z)}  G_0(z,z')
\label{eq:green}
\end{equation}
and $G_0$ is the Coulomb green function
\begin{equation}
-\partial^2 G_0(z,z')= \delta^{(3)}(z-z')
\label{eq:greencoulomb}
\end{equation}
By direct substitution of eq~(\ref{eq:greencoulomb}) into eq~(\ref{eq:green}),
we can write the exponent $I[ j_{\mu}^{\beta}  ]$ in the form
\begin{equation}
I [ j_{\mu}^{\beta} ]= \sum _{\alpha\beta}\;
{ {({\kappa }^{-1})^{\alpha\beta}} \over 2}
 \int dz\int dz' \;j_{\mu}^{\alpha}(z) \epsilon^{\mu \nu \lambda}
\partial_{\lambda}^{(z)} G_0(z,z') \; j_{\nu}^{\beta}(z')
\label{eq:iii}
\end{equation}

Using the magnetostatic analogy, we can now regard $j_{\mu}^{\alpha}$ as a
current in three dimensional Euclidean
space and use it to evaluate the expressions in eq~(\ref{eq:iii}).
Let ${\cal C}_{\mu}^{\alpha}$ be a vector field
related to $j_{\mu}^{\alpha}$ by the equation (``Amp\`ere's Law")
\begin{equation}
{\vec {\bigtriangledown}} \times \vec {\cal C}^{\alpha}= {\vec j}^{\alpha}
\label{eq:amp}
\end{equation}
such that
\begin{equation}
{\vec {\bigtriangledown}} \cdot {\vec {\cal C}}^{\alpha}=0
\label{eq:rot}
\end{equation}
The solution of eq~(\ref{eq:amp}), subject to the constraint eq~(\ref{eq:rot}),
can be found in the same way as in reference \cite{l1}.
The field ${\cal C}_{\mu}^{\alpha}$ is given by
\begin{equation}
{\cal C}_{\nu}^{\alpha}(z)=\int d^3w \epsilon_{\nu \lambda \mu}
            \partial_{\lambda}^{(z)} \; G_0(z,w) j_{\mu}^{\alpha}(w)
\label{eq:cur}
\end{equation}
By substituting eq~(\ref{eq:cur}) back into eq~(\ref{eq:iii}), we find that
$I [j_{\mu}^{\beta} ]$ takes the simpler form
\begin{equation}
I [ j_{\mu}^{\beta} ]= \sum _{\alpha\beta}\;
{ {({\kappa }^{-1})^{\alpha\beta}} \over 2}
 \int d^3 z \; {\cal C}_{\mu}^{\alpha}(z) j_{\mu}^{\alpha}(z)
\label{eq:ij}
\end{equation}
Now, since the currents $j_{\mu}^{\alpha}$ are non-zero only on the world
lines, we
can rewrite the volume integral in eq~(\ref{eq:ij}) in the form of a line
integral
over the configuration $\Gamma^{\alpha}$. The set of closed loops
$\Gamma ^{\alpha}$ are the boundary a surface $\Sigma^{\alpha}$,
$\Gamma ^{\alpha}=\partial \Sigma ^{\alpha}$. We can then apply
Stokes' theorem to get the result
\begin{equation}
I [ j_{\mu}^{\beta}] = \sum _{\alpha\beta}\;
{ {({\kappa }^{-1})^{\alpha\beta}} \over 2}
 \int_{{\Sigma}^{\alpha}} d \sigma \; n_{\mu}^{\alpha} \;j_{\mu}^{\beta}
\label{eq:ifin}
\end{equation}
where $n_{\mu}^{\alpha}$ is a vector field normal to the surface
$\Sigma^{\alpha}$.
The integral
$\int_{{\Sigma}^{\alpha}} d \sigma \; n_{\mu}^{\alpha} \; j_{\mu}^{\beta}$
is an integer which counts the number of times the current $j_{\mu}^{\beta}$
pierces the surface $\Sigma ^{\alpha}$.

The result of eq~(\ref{eq:ifin}) means that the
average over the statistical gauge fields eq~(\ref{eq:pi}) has the simple form
\begin{equation}
\langle \exp (i \sum _{\alpha}\;\int d^3z j_{\mu}^{\alpha}(z)
A^{\mu}_{\alpha }(z))\rangle_{\rm CS}= \exp{(i \sum _{\alpha\beta}\;
{ {({\kappa }^{-1})^{\alpha\beta}} \over 2}
\int_{{\Sigma}^{\alpha}} d \sigma \; n_{\mu}^{\alpha} \; j_{\mu}^{\beta})}
\label{eq:refin}
\end{equation}
We are now in position to ask when are the two systems equivalent. In other
words, when is the average over the statistical fields equal to one or, at
least, independent of the integral in eq~ (\ref{eq:ifin}). In the first case
we would have proven that the amplitude in the system with the statistical
gauge fields is exactly equal to the same amplitude calculated in their
absence, whereas in the second case all amplitudes differ by a constant phase
factor, which can be dropped. By inspecting eq~(\ref{eq:refin}) we see that the
amplitude is equal to one if
\begin{eqnarray}
\kappa^{-1}= 2 \pi \left(
\begin{array}{cc}
2 s_1  & n \\
n & 2 s_2
\end{array}
\right)
\label{eq:Kmenos}
\end{eqnarray}
where $s_{1}$,$s_{2}$, and $n$ are arbitrary integers.

\section{}
\label{sec:BB}

In this Appendix, we derive an expression for the energy of the ground state
which includes the corrections due to the gaussian fluctuations of the
Chern-Simons gauge fields.
We evaluate this energy for two different incompressible states with filling
fraction $\nu =1$.We show that the mean field degeneracy is lifted.
In particular, within our approximation, we show that the ($1,1,1$) state
is the lowest energy state.

We begin with the derivation of an expression for the ground state energy.
The quantum partition function for this problem at zero temperature is
given by eq~(\ref{eq:pf}). In order to calculate the ground state energy,
we will set the external electromagnetic field to zero.
Since the action is quadratic in the fermions, they can be integrated out.
After expanding the Chern-Simons gauge fields around their mean field
solution $<a_{\mu}^{\alpha}>$, the partition function can be written as follows
\begin{equation}
{\cal Z} =  e^{iS_{\rm MF}}
   \int {\cal D} {\tilde a}_{\mu}^{\alpha} \;
e^{i S_{\rm eff}({\tilde a}_{\mu}^{\alpha})}
\label{eq:pfapp}
\end{equation}
where $S_{\rm MF}$ coincides with the expression for the effective action
in eq~(\ref{eq:ese2}) but setting $a_{\mu}^{\alpha}=<a_{\mu}^{\alpha}>$
and ${\tilde A}_{\mu}=0$, and $S_{\rm eff}({\tilde a}_{\mu}^{\alpha})$ is
given by eq~(\ref{eq:gauss}) with ${\tilde A}_{\mu}=0$.
 If we define ${\bar \Pi}^{\mu \nu}_{\alpha\beta}(x,y)$ as the sum of the
polarization tensor $\Pi^{\mu \nu}_{\alpha\beta}(x,y)$ plus the tensors
corresponding to the interaction term and the Chern-Simons term in
eq~(\ref{eq:gauss}), we can write the effective action in eq~(\ref{eq:pfapp})
as
\begin{equation}
S_{\rm eff}({\tilde a}_{\mu}^{\alpha})=
      {1\over 2}  \int d^{3}x \int d^{3}y \; {\tilde a}_{\mu}^{\alpha}(x)\;
                  {\bar \Pi}^{\mu \nu}_{\alpha\beta}(x,y) \;
                {\tilde a}_{\nu}^{\beta}(y)
\label{eq:app1}
\end{equation}
Since this action is quadratic in the fluctuations of the Chern-Simons gauge
fields, they can be integrated out explicitly. The partition
function becomes
\begin{equation}
{\cal Z}= e^{iS_{\rm MF}}[{\cal D}{\rm et}({\bar \Pi}^{\mu \nu}_{\alpha\beta})]
^{-{1\over 2}}
\label{eq:pfapp2}
\end{equation}
where ${\cal D}{\rm et}({\bar \Pi}^{\mu \nu}_{\alpha\beta})$ is the
functional determinant of the gaussian fluctuation operator. Thus, we
can write ${\cal Z}= e^{i S_{\rm tot}}$ where \begin{eqnarray}
S_{\rm tot}&=& {S_{\rm MF}}+ {i\over 2}
\ln  {\cal D}{\rm et}({\bar \Pi}^{\mu \nu}_{\alpha\beta}) \nonumber \\
        &=&{S_{\rm MF}}+ {i\over 2}
{\rm Tr} \ln  ({\bar \Pi}^{\mu \nu}_{\alpha\beta}) \nonumber \\
        &=&{S_{\rm MF}}+ {i\over 2}
\int {d^2Q \over (2\pi)^2} \int {d \omega \over (2\pi)}
{\rm tr} \ln  ({\bar \Pi}^{\mu \nu}_{\alpha\beta}) \nonumber \\
        &=&{S_{\rm MF}}+ {i\over 2}
\int {d^2Q \over (2\pi)^2} \int {d \omega \over (2\pi)}
\ln  \det ({\bar \Pi}^{\mu \nu}_{\alpha\beta})
\label{eq:stot}
\end{eqnarray}
In eq.~(\ref{eq:stot}) ${\rm Tr} {\hat A}$ stands for the functional
trace of an operator $\hat A$ while ${\rm tr} M$ is the algebraic trace
of a finite matrix $M$ over the indices $\alpha, \beta, \mu$ and
$\nu$. Similarly, $\det
M$ is the algebraic determinant of the matrix $M$.

In the limit of infinite time span, $T
\rightarrow \infty$ (or zero temperature), the quantum partition
function
can be written in terms of the ground state energy $E_{\rm GS}$, as
${\cal Z}= e^{-T L^{2} E_{\rm GS}}$,
where $L^2$ is the area of the system.
 Therefore, using eq~(\ref{eq:stot}), we find that the ground state energy
is given by
\begin{equation}
E_{\rm GS}= E_{\rm GS}^{\rm MF} - {i\over {2TL^2}}
\int {d^2Q \over (2\pi)^2} \int {d \omega \over (2\pi)}
 \ln(\det {\bar \Pi}^{\mu \nu}_{\alpha\beta})
\label{eq:pfapp6}
\end{equation}
It can be shown that $ \det{\bar \Pi}^{\mu \nu}_{\alpha\beta}$ is equal
to the square of the denominator of the density correlation function,
$ D(\omega, {\vec Q})$, {\it i.e.},
$ \det{\bar \Pi}^{\mu \nu}_{\alpha\beta} = \left[ D(\omega, {\vec
Q})\right]^2$. Here, as in the single
layer case, the zeroes of the denominator of the density correlation
function determine their poles. Therefore, the zeroes of
$D(\omega, {\vec Q})$ are the collective modes of the system. In other words,
we can write
$D(\omega, {\vec Q})= {\prod_{\lambda}} ({\omega}^2
-{\omega}^2_{\lambda})$,
where ${\omega}^2_{\lambda}$ are the collective modes of the system.

To evaluate the integral in the r.h.s of eq~(\ref{eq:pfapp6}) we
consider a cutoff function $f_{\Lambda}(\omega, {\vec Q})$ such that
$f_{\Lambda}(\omega, {\vec Q}) \rightarrow 0$ when
$\omega \rightarrow \pm \infty$, and
$\partial_{\omega} f_{\Lambda}(\omega, {\vec Q}) \rightarrow 0$,{\it i.e.},
it is an adiabatic cutoff. Therefore,
the integral in the r.h.s of eq~(\ref{eq:pfapp6}) can be calculated as
follows
\begin{eqnarray}
\int {d \omega \over (2\pi)}
\ln  \det ({\bar \Pi}^{\mu \nu}_{\alpha\beta})
& = & 2
\int {d \omega \over (2\pi)}  f_{\Lambda}(\omega, {\vec Q})
\ln   D(\omega, {\vec Q}) \nonumber \\
&= & -2 \int {d \omega \over (2\pi)}  f_{\Lambda}(\omega, {\vec Q})
{ \omega \over D(\omega, {\vec Q})}{\partial _{\omega}}D(\omega, {\vec Q})
\label{eq:rhs}
\end{eqnarray}
where the last equality was obtained integrating by parts.
Using this result eq~(\ref{eq:pfapp6}) becomes
\begin{equation}
E_{\rm GS}- E_{\rm GS}^{\rm MF}= {i\over {TL^2}}
\int {d^2Q \over (2\pi)^2} \int {d \omega \over (2\pi)}
f_{\Lambda}(\omega, {\vec Q})
{ \omega \over D(\omega, {\vec Q})}{\partial _{\omega}}D(\omega, {\vec Q})
\label{eq:pfapp7}
\end{equation}
After integrating in $\omega$ we get
\begin{equation}
E_{\rm GS}- E_{\rm GS}^{\rm MF}= {1\over {TL^2}}
\int {d^2Q \over (2\pi)^2} \sum_{\lambda}
{\omega}^2_{\lambda} f_{\Lambda}({\vec Q})
\label{eq:pfapp8}
\end{equation}
In eq.~(\ref{eq:pfapp8}) the sum over modes includes the case of
eventual multiplicities and degeneracies.

We will evaluate this expression for the state I, which has $\nu=1$ and
$p_1=p_2=1$, $s_1=s_2=0$ and $n=1$, and for the state II, which has also
$\nu=1$ and $p_1=p_2=1$, but $s_1=s_2=1$ and $n=-1$. It can be shown that the
collective excitations of the state II coincide with the ones of the
state ($3,3,1$) (derived in Section   \ref{subsec:unmedio}), but replacing
${v}_{12}$ by $-{v}_{12}$, and using that
${\bar \omega}={{\omega_c} \over 2}$ and ${\bar B}= {B \over 2}$.
Substituting the expressions for the collective modes calculated in Section
\ref{subsec:ememe} for the state I, and the ones calculated in Section
\ref{subsec:unmedio} for the state II, (but replacing
${v}_{12}$ by $-{v}_{12}$, ${\bar \omega}={{\omega_c} \over 2}$ and
${\bar B}= {B \over 2}$), into eq~(\ref{eq:pfapp8}), we obtain
\begin{equation}
E_{\rm GS}^I- E_{\rm GS}^{II}= {{{\bar \omega}^2 {\bar B}} \over {2\pi}}
{1\over {TL^2}}
\int_0^{\infty} {dx} [-4-(1+{\tilde M}) {x\over 2} + (2x(1+{\tilde
M}))^{1\over 2}] f_{\Lambda}(x)
\label{eq:pfapp9}
\end{equation}
where ${\tilde M}={M\over {2\pi}}(v_{11} - v_{12})$, $x={{\vec Q}^2 \over
{2{\bar B}}}$, and we have kept only linear terms in $x$.
In particular, using a sharp cutoff to calculate the integral in the r.h.s of
eq~(\ref{eq:pfapp9}), such as
${\vec Q}^2 < 2{\bar B}$, we find that $E_{\rm GS}^I- E_{\rm GS}^{II} <0$, in
both cases, for Coulomb and for short range interactions.

A few comments on this result are in order. Firstly, since this theory
does not have a small parameter, it may be argued that a result obtained
in the leading, gaussian, correction to the mean field result might be
affected significantly by higher order corrections. For a generic pair
interaction, this should be the case and the leading order result may
not be so significant. Notice,
however, that the approximations that we have made are accurate only for
interactions whose characteristic length scales are smaller than the
cyclotron radius. In this situation, the gaussian effects should become
dominant and the selected state should be unique. However,
if the interaction is important only at
distances long compared with the cyclotron radius, the selection of the
states may be different and a complex phase diagram should then be
expected. But, in such cases,
other states , such as crystals, may also compete effectively with the
fluid states.

Thus, we have shown that, for an interaction with a sufficiently short
range, the state $(1,1,1)$ is the fluid ground state. We believe that
similar arguments can be made in all the other cases where we found
multiple solutions (with the same caveats about the range of the
interactions).

\end{document}